\documentclass[twocolumn]{aastex631}
\usepackage{graphicx,float,subfigure} 
\usepackage{booktabs}
\usepackage{float}
\usepackage[T1]{fontenc}
\usepackage[utf8]{inputenc}
\usepackage{ae}
\usepackage{aecompl}
\usepackage{amsmath, amssymb, amsfonts, braket, gensymb} 
\usepackage{newtxtext,newtxmath} 
\usepackage{enumitem}
\usepackage{longtable}
\usepackage{hyperref}
\usepackage{spverbatim}
\hypersetup{
    colorlinks=true,
    linkcolor=blue,
    filecolor=magenta,
    pdftitle={Overleaf Example},
    pdfpagemode=FullScreen,
    }
\urldef{\urlA}\url{https://isochrones.readthedocs.io/en/latest/starmodel.html#Priors} 

\newcommand{\code}[1]{\texttt{#1}}

\shorttitle{Accurate, Precise, and Physically Self-consistent Subgiant Ages}
\shortauthors{Nataf et al.}

\begin{document}

\title{Accurate, Precise, and Physically Self-consistent Ages and
Metallicities for 400,000 Solar Neighborhood Subgiant Branch Stars}

\correspondingauthor{David M. Nataf}
\email{dnataf1@jhu.edu, david.nataf@gmail.com}

\author[0000-0001-5825-4431]{David M. Nataf}
\affiliation{Center for Astrophysical Sciences, Johns Hopkins University,
3400 N Charles St, Baltimore, MD 21218, USA}
\affiliation{William H.\ Miller III Department of Physics \& Astronomy,
  Johns Hopkins University, 3400 N Charles St, Baltimore, MD 21218, USA}
\affiliation{Department of Physics \& Astronomy, University of Iowa, Iowa City, IA 52242, USA}

\author[0000-0001-5761-6779]{Kevin C.\ Schlaufman}
\affiliation{William H.\ Miller III Department of Physics \& Astronomy,
Johns Hopkins University, 3400 N Charles St, Baltimore, MD 21218, USA}

\author[0000-0001-6533-6179]{Henrique Reggiani}[]
\affiliation{Gemini Observatory/NSF's NOIRLab, Casilla 603, La Serena, Chile}
\affiliation{The Observatories of the Carnegie Institution for Science,
813 Santa Barbara St, Pasadena, CA 91101, USA}

\author[0000-0003-4580-9893]{Isabel Hahn}
\affiliation{The Observatories of the Carnegie Institution for Science,
813 Santa Barbara St, Pasadena, CA 91101, USA}
\affiliation{Department of Physics and Astronomy, Pomona College,
Claremont, CA 91711, USA}

\begin{abstract}

\noindent
Age is the most difficult fundamental stellar parameter to infer for
isolated stars.  While isochrone-based ages are in general imprecise
for both main sequence dwarfs and red giants, precise isochrone-based
ages can be obtained for stars on the subgiant branch transitioning
from core to shell hydrogen burning.  We synthesize Gaia DR3-based
distance inferences, multiwavelength photometry from the ultraviolet to
the mid infrared, and three-dimensional extinction maps to construct a
sample of 289,759 solar-metallicity stars amenable to accurate, precise,
and physically self-consistent age inferences.  Using subgiants in the
solar-metallicity open clusters NGC 2682 (i.e., M  67) and NGC 188, we
show that our approach yields accurate and physically self-consistent
ages and metallicities with median statistical precisions of 8\%
and 0.06 dex.  The inclusion of systematic uncertainties resulting
from non-single or variable stars results in age and metallicity
precisions of 9\% and 0.12 dex.  We supplement this solar-metallicity
sample with an additional 112,062 metal-poor subgiants, including over
3,000 stars with $[\text{Fe/H}]\lesssim-1.50$, 7\% age precisions, and
apparent Gaia $G$-band magnitudes $G<14$.  We further demonstrate that
our inferred metallicities agree with those produced by multiplexed
spectroscopic surveys.  As an example of the scientific potential
of this catalog, we show that the solar neighborhood star-formation
history has three components at $([\text{Fe/H}],\tau/\text{Gyr}) \approx
(+0.0,4)$, $(+0.2,7)$, and a roughly linear sequence in age--metallicity
space beginning at $([\text{Fe/H}],\tau/\text{Gyr})\approx(+0.2,7)$
and extending to $(-0.5,13)$.  Our analyses indicate that the solar
neighborhood includes stars on disk-like orbits even at the oldest ages
and lowest metallicities accessible by our samples.

\end{abstract}


\section{Introduction} \label{sec:Introduction}

Inferring the ages of isolated field stars, and by extension, the
star-formation history of the solar neighborhood, remains among the most
challenging and most fundamental problems of Galactic astronomy (e.g.
\citealt{1980ApJ...242..242T,2010ARA&A..48..581S,2016MNRAS.455..987C}).
Several approaches are currently being both widely-used
and developed, including asteroseismic age estimates
\citep{2019MNRAS.490.5335S,2021A&A...645A..85M},
probabilisticaly-weighted isochrone-based ages
\citep{2004MNRAS.351..487P,2005A&A...436..127J,2018MNRAS.477.2326F,2021AJ....161..147B},
rotation-based ages \citep{2007ApJ...669.1167B,2012ApJ...746..102C},
and radioactivity-based ages
\citep{1996ApJ...467..819S,2002ApJ...572..861C}. Each of these comes
with their own specific strengths, successes, selection effects,
and limitations.

In this investigation we focus on a specific class of isochrone-based
ages -- that of stars along the subgiant branch. These are the stars
which are transitioning from core to shell hydrogen burning. The
luminosities of subgiants are strong functions of their core masses
\citep{2005essp.book.....S,2013sse..book.....K}. Since core mass scales
with stellar mass, and main-sequence lifetime scales with stellar
mass, luminosity on the subgiant branch is also a function of age.
It can be estimated quantitatively (see Figure \ref{fig:selection})
that the absolute magnitude $M$ of a subgiant star scales with age at
a level approximating:
\begin{equation}
 M_{1} - M_{2} \approx  \ln ({\tau_{2}} / {\tau_{1}}),
 \label{EQ:precision}
\end{equation}
and thus a 1\% measurement in the absolute magnitude of a subgiant star
approximately corresponds to a 1\% measurement in its \textit{relative,
model-dependent} age, with the exact dependence being a function
of location on the subgiant branch, age, and metallicity. It
follows that if parallax measurements with 1\% precision are available,
the theoretical lower bound on the derived ages of subgiants is
approximately 2\%. Among the other advantages of subgiants: i)
They are bright, and thus detailed elemental abundance measurements
can be obtained to large distances, and ii) they are numerous for
stellar populations older than $\tau \approx 1$~Gyr, enabling study
of intermediate and old stellar populations with a quantifiable
bias.

In this investigation we neglect to account for possible
variations in the initial helium abundances of stars, of their initial rotation, which can affect
the relationship between ages and luminosity on the subgiant branch
\citep{2010ApJ...714.1072M,2012A&A...547A...5V,2012AcA....62...33N,2015A&A...577A..72V}. A recent investigation of subgiants with Kepler asteroseismology has found that age inferences of subgiant stars are much more sensitive to uncertainties in the assumed metallicities than in either the initial helium abundance or the mixing length assumed by stellar models \citep{2020MNRAS.495.3431L}. It is also the case that prior investigations have shown that scaled-solar helium abundances are a good fit to the data for most solar neighbourhood stars \citep{2007MNRAS.382.1516C}.  Unidentified blends and binaries are undoubtedly a source of systematic error, which we explore later in this work.

A recent study of the local stellar age distribution using $\sim$250,000
subgiant stars is that of \cite{2022Natur.603..599X}. Their
methodology differs from ours in several respects, most
significantly that they constrain their stellar metallicities
using spectroscopically-derived measurements from LAMOST
\citep{2012RAA....12..735D,2012RAA....12..723Z,2019ApJS..245...34X},
whereas we use photometric measurements across the available
wavelength regime, including ultraviolet measurements which are the
most sensitive to metallicity. We also note that in principle
one can fit for stellar parameters to the spectra themselves, as done
by \citet{2014MNRAS.443..698S}, rather than to the stellar parameters
derived from the spectra.

The advantage of the approach that we develop and employ here
is the potential for a large, all-sky sample of subgiants
with exquisitely-measured stellar parameters. Much of our
ultraviolet data comes from GALEX \citep{2007ApJS..173..682M},
which will be augmented in the future with missions
such as ULTRASAT \citep{2023arXiv230414482S} and UVEX
\citep{2021arXiv211115608K}. The significant leverage of GALEX
photometry in inferring stellar atmospheric parameters such as metallicity
\citep{2014AstBu..69..160S,2019ApJ...872...95M,2023arXiv231116901L},
in combination with all-sky astrometric, photometric,
variability, and spectroscopic surveys such as Gaia
\citep{2016A&A...595A...1G}, as well as all-sky extinction maps
\citep{2017A&A...606A..65C,2019ApJ...887...93G,2022A&A...664A.174V},
we should eventually be able to measure the precise stellar parameters
(including age) for up to  $ \mathcal{O}(10^7) $ subgiants. That is a
regime where we can expect to be able to resolve individual star-formation
events, and thus bring unprecedented resolution to the story of the
Milky Way's formation and assembly.

In this investigation, we develop the method of astro-photometric age
and composition inferences of subgiant stars, with the aims of each of
producing a catalog for further study, identifying and quantifying the
strengths and limitations of this method, and charting a path for robust
future study. The structure of this paper is as follows. In Section
\ref{sec:SampleSelection}, we describe how we build our sample. In
Section \ref{sec:Clusters}, we test our methodology in seven specific
ways. Our results are presented in Section \ref{sec:Results}. We conclude
in Section \ref{sec:Conclusion}.

\section{Sample Selection}  \label{sec:SampleSelection}

\begin{figure}
\includegraphics[width=0.48\textwidth]{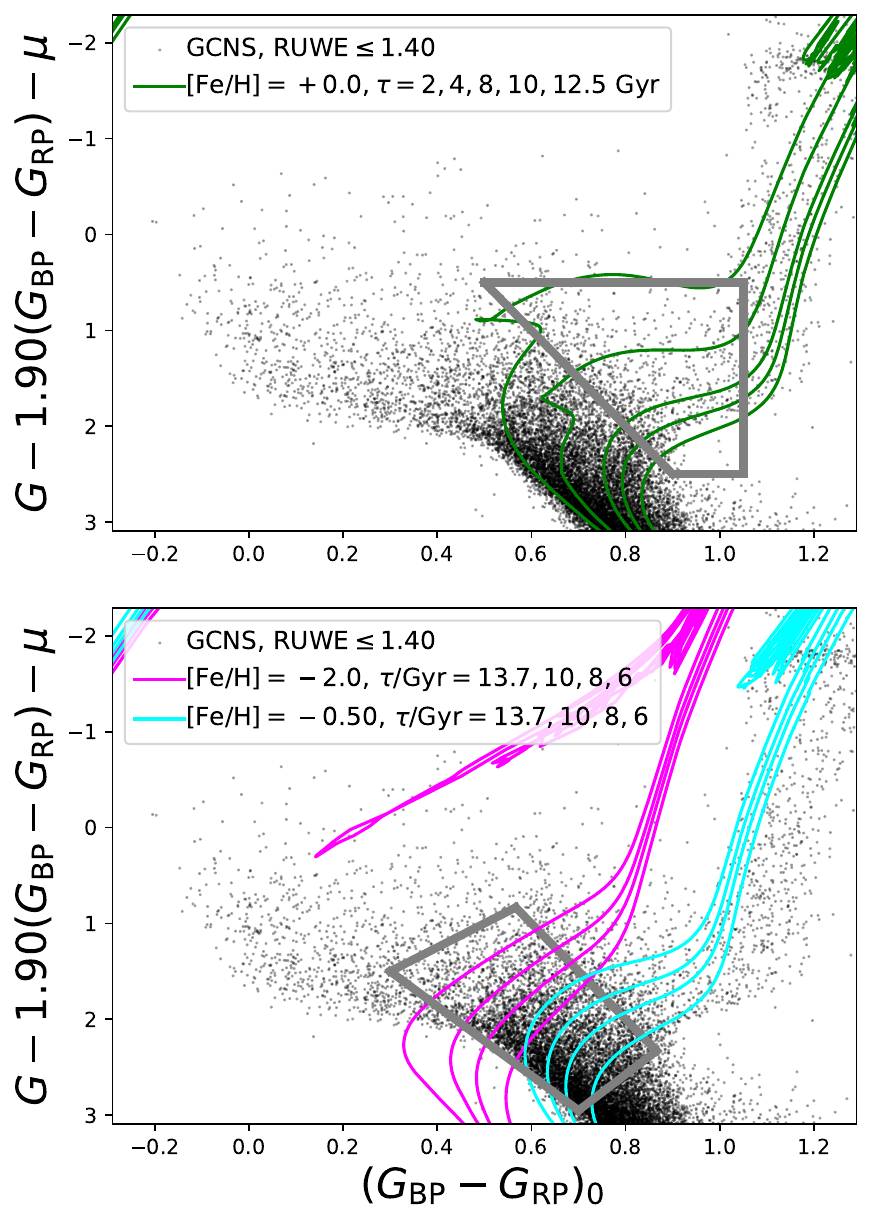}
\caption{We show our color-magnitude selection function (thick grey
lines) overplotted on the Gaia Catalogue of Nearby Stars (small black
points) with MIST isochrones for the main sample (TOP) and the metal-poor
annex (BOTTOM). }
\label{fig:selection}
\end{figure}

We select our subgiant sample using a combination of Gaia DR3 magnitudes,
distances from \citet{2021AJ....161..147B}, and reddening estimates as
discussed in Section \ref{subsec:data} and in our appendix. The
magnitudes that we use for our sample selection are the nearly
extinction-independent absolute Wesenheit magnitudes,
\begin{equation}
W_{G}=G - 1.90(G_{\rm{BP}}-G_{\rm{RP}})-5\log(d/pc)+5,
\end{equation}
and colors that we use are:
\begin{equation}
(G_{\rm{BP}}-G_{\rm{RP}})_{0} = (G_{\rm{BP}}-G_{\rm{RP}})-0.424A_{V}.
\end{equation}
The color and magnitude criteria for our main sample (top panel of Figure
\ref{fig:selection}) are:
\begin{itemize} [itemsep=0.0cm]
    \item $W_{G} \leq 2.50$;
    \item $W_{G} \geq +0.50$;
    \item $W_{G} \leq 2.50+5((G_{\rm{BP}}-G_{\rm{RP}})_{0}-0.90)$;
    \item $(G_{\rm{BP}}-G_{\rm{RP}})_{0} \leq 1.05$;
    \item At least one measurement of GALEX $NUV$, Skymapper $u$, or
    SDSS $u$, hereafter denoted $NUV$, $u_{\rm{SM}}$, $u_{\rm{SDSS}}$.
\end{itemize}
This subgiant sample inevitably includes some stars on the turnoff
and the base of the red giant branch, as the location of these phases
of stellar evolution on the color-magnitude diagram is a sensitive
function of age and metallicity. The primary sample defined above is most
efficient at targeting stars with $-0.50 \lesssim \rm{[Fe/H]} \lesssim
+0.50$. Subgiant stars of lower metallicity appear at the same location
in the Gaia DR3 color-magnitude diagram as the vastly more numerous
more metal-rich turnoff stars, and thus finding them would otherwise
be akin to searching for a needle in a haystack, but we can delineate
them by making use of metallicity-sensitive ultraviolet photometry. We
select the metal-poor annex (bottom panel of Figure \ref{fig:selection})
using the following criteria:
\begin{itemize} [itemsep=0.0cm]
    \item $W_{G} > 2.50+5((G_{\rm{BP}}-G_{\rm{RP}})_{0}-0.90)$;
    \item $W_{G} \geq 1.5 - 2.5((G_{\rm{BP}}-G_{\rm{RP}})_{0}-0.30)$;
    \item $W_{G} \leq 1.5 + 3.625((G_{\rm{BP}}-G_{\rm{RP}})_{0}-0.30)$;
    \item $W_{G} \leq 2.95-3.75((G_{\rm{BP}}-G_{\rm{RP}})_{0}-0.70)$;
    \item At least one measurement of GALEX $NUV$, Skymapper $u$, or SDSS $u$;
    \item Each available ultraviolet measurement satisfies the
    relevant of: $(NUV-G) \leq -0.65 + 13/2(G_{\rm{BP}}-G_{\rm{RP}})$,
    $(u_{\rm{SM}}-G) \leq 0.19 + 11/6(G_{\rm{BP}}-G_{\rm{RP}})$, and
    $(u_{\rm{SDSS}}-G) \leq 0.11 + 11/6(G_{\rm{BP}}-G_{\rm{RP}})$.
\end{itemize}
We also require the following photometric, variability, reliability,
and reddening criteria, for both samples:
\begin{itemize} [itemsep=0.0cm]
    \item Measurements in all three of of $J$, $H$, $K_{s}$ from 2MASS;
    \item Measurements in both of $W_{1}$ and $W_{2}$ from WISE;
    \item From Gaia DR3, \texttt{parallax\_over\_error $\geq$ 50};
    \item From Gaia DR3, \texttt{duplicated\_source $=$ 'FALSE'};
    \item From Gaia DR3, \texttt{phot\_proc\_mode $=$ 0};
    \item From Gaia DR3, \texttt{phot\_variable\_flag $!=$ 'VARIABLE'};
    \item From Gaia DR3, \texttt{non\_single\_star $=$ 0};
    \item From Gaia DR3, \texttt{ruwe $\leq$ 1.4};
    \item From Gaia DR3, \texttt{ipd\_gof\_harmonic\_amplitude $\leq$ 0.10};
    \item From Gaia DR3, \texttt{astrometric\_params\_solved $=$ 31}  ;
    \item No match in the ASAS-SN catalog of variable stars
    \citep{2014ApJ...788...48S,2021MNRAS.503..200J,2022arXiv220502239C};
    \item $A_{V} \leq 0.50$.
\end{itemize}
The selection function (in Gaia photometry) is shown in Figure
\ref{fig:selection}, where we also show the color-magnitude diagram
of the Gaia Catalogue of Nearby Stars \citep{2021A&A...649A...6G}
and [Fe/H]$=-2.00$, $-0.50$, and $+0.00$ isochrones from MIST
\citep{2011ApJS..192....3P,2013ApJS..208....4P,2015ApJS..220...15P,2016ApJS..222....8D,2016ApJ...823..102C}
for comparison.

The selection function described in this section formed from a combination
of guidelines in the input data, as well as arbitrary delineations meant
to optimize the trade-off between selecting more subgiants and selecting
fewer turnoff and red giant stars.

\subsection{Additional Photometric Criteria}
\label{subsec:MorePhot}

Ideally, cross-matching point sources between different photometric
surveys would be conducted using purely astrometric criteria. In practice,
that leads to spurious cross-matches due to factors such as the varying
astrometric precision and accuracy between the different surveys,
the varying size of the point-spread function and thus sensitivity to
blends, that the sensitivity to blends will itself be wavelength and thus
bandpass-dependent, and the different saturation limits and photometric
depths of each survey.

We implement the \textit{ad hoc} requirement that in the posterior
calculation for each star, the predicted apparent magnitudes in GALEX
$NUV$, 2MASS $J$, WISE $W_{2}$, SDSS $ug$, Skymapper $uvg$, and Pan-STARRS
$gy$ differ by no more than 3 sigma with the observed magnitudes. In
each case, the photometric errors given by the catalogs are inflated by
0.01 mag in quadrature, and the total number of stars flagged as having
less certain photometric matches is 9,609 in the primary sample. We
find that the number of flagged stars is typically 5$\times$ what one
would expect from purely statistical errors. Upon inspection, we find
that most of the matches are due to large offsets between apparent and
predicted magnitudes -- plausibly mismatches or blends.

In the example of $NUV$,our criterion flags 2,936 stars from our sample
of 229,156 stars with $NUV$ matches, which is approximately 5$\times$
greater than the expectation from pure statistical expectations. For
those flagged stars, the mean and 1-$\sigma$ astrometric offset between
Gaia DR3 and GALEX is $(0.7 \pm 0.5) \arcsec$, whereas for the sample as
a whole the separation is distributed as $(0.5 \pm 0.4) \arcsec$. The
larger mean and scatter in the astrometric separations are consistent
with the inclusion of more spurious matches.

\subsection{Additional Variability Criteria}
\label{subsec:MoreVar}

We use two additional variability criteria to flag stars as less reliable
probes of single-star stellar evolution. We still compute the posteriors
for these parameters of these stars, and still report them in our data
tables, but we do not include the results in our analysis.

The first additional criterion is based on the Gaia G-band photometric
variability of the star. For each star in our main sample, we compute
the photometric variability $\sigma_{G}$ as follows:
\begin{equation}
\sigma_{G} = (\rm{phot\_g\_mean\_flux\_error}) (\sqrt{phot\_g\_n\_obs}),
\end{equation}
and we then compute the percentile for $\sigma_{G}$ for each star at
the apparent magnitude of that star, in our primary sample. We use
that percentile as an indicator of photometric variability, though it
is certainly a coarse proxy, as the number of measurements are sparse
for many of the stars.

We find that that requiring that the percentile of $\sigma_{G}$ be below
95th yields a reasonable metric for removing variables. For stars with
$\sigma_{G}$ below the 95th percentile, approximately 4\% have derived
[Fe/H] less than or equal to $-0.50$, whereas the fraction rises to 9\%
for stars with $\sigma_{G}$ above the 95th percentile.  In all cases
discussed here, we also remove stars flagged as having less reliable
photometric matches. A reasonable explanation for this trend, whereby
more photometrically variables are inferred to more frequently have a
low metallicity, is that rapidly rotating stars have bluer spectra,
with the effect becoming larger at smaller wavelengths, mimicking
metal-poor atmospheres (Luca Casagrande, private communication, using
models discussed in \citealt{2014MNRAS.444..392C}).

The second criterion makes use of the work of \citet{2022arXiv220611275C},
who studied the inferred measurement errors in the radial velocity
measurements reported by Gaia. They showed that these are associated
with spectroscopic binaries, which have the effect of broadening the
spectra that are binned from different measurements taken at different
times. In the context of our study, unresolved spectroscopic binaries
would shift the colors from those expected from single-stellar evolution,
and also include a population of past or present mass-transfer binaries,
such as blue stragglers.

The criterion suggested by \citet{2022arXiv220611275C} is one where the
p-value to the radial velocity noise satisfies $p \leq 0.001$, where they
stress that the p-value in their analysis represents the likelihood that
the source’s radial velocity noise is indeed anomalously high compared
to stars of similar color and magnitude.

Of the 289,756 stars in our primary subgiant sample, 266,443 are not
flagged as either having less reliable photometric crossmatches or
having photometric variability above the 95th percentile. Of those,
\citet{2022arXiv220611275C} reported p-values for 118,636 Gaia sources,
for which 23,898, or 20\%, have p-values $\leq$ 0.001 and thus are likely
spectroscopic binaries. These 23,898 stars are flagged in our data tables
and removed from subsequent analysis discussed in this paper. Our total
sample used for analysis in this paper are those stars with variability
below the 95th percentile, with no photometry flagged as spurious, and
for which \citet{2022arXiv220611275C} either reported no p-value or one
greater than 0.001, this sample numbers 242,545 point sources.

\subsection{Selection of Astrometric, Photometric, and Extinction data}
\label{subsec:data}

Our analysis incorporates the following data as input:
\begin{enumerate} [itemsep=0.0cm]
\item Parallaxes ($\pi$) measurements and associated uncertainties
from Gaia Early Data Release 3 \citep{2023AnA...674A...1G},
as well as estimates on the distances derived therefrom
\citep{2021AJ....161..147B} and parallax zero-point corrections from
\citet{2021A&A...649A...4L}. $G_{\rm{BP}}$, $G$, and $G_{\rm{RP}}$
magnitudes and associated uncertainties are taken from Gaia Data Release
2 \citep{2018A&A...616A...1G}.
\item $FUV$ and $NUV$ magnitudes and associated uncertainties
are taken from the revised catalog of GALEX ultraviolet sources
\citep{2017ApJS..230...24B}.
\item $u$, $g$, $r$, $i$, and $z$ magnitudes and associated uncertainties
are taken from the Sloan Digital Sky Survey \citep{2020ApJS..249....3A}.
\item $u$, $v$, $g$, $r$, $i$, and $z$ magnitudes and associated
uncertainties are taken from Data Release (DR) 2 of the SkyMapper Southern
Sky Survey \citep{2019PASA...36...33O}.
\item $g$, $r$, $i$, $z$, $y$ photometry are taken from the Pan-STARRS1
catalog \citep{2020ApJS..251....7F}.
\item $J$, $H$, and $K_{\text{s}}$ magnitudes and associated uncertainties
are taken from the 2MASS Point Source Catalog \citep{2006AJ....131.1163S}.
\item $W1$ and $W2$ magnitudes and associated uncertainties are taken
from the Wide-field Infrared Survey Explorer (WISE) AllWISE Source Catalog
\citep{2010AJ....140.1868W,2011ApJ...743..156M,2014yCat.2328....0C}.
\item We use the extinction maps of \citet{2019ApJ...887...93G},
setting $A_{V}=3.04 E(g-r)$, for those sightlines for which the
quality flags for \texttt{'converged'} and \texttt{'reliable\_dist'}
are both equal to \texttt{'TRUE'}. For the majority of
the remaining sightlines, we use the extinction estimates of
\citet{2022A&A...661A.147L}/\citet{2022A&A...664A.174V}, specifically
from the version 2 medium-resolution 6 kpc $\times$ 6 kpc $\times$ 0.8
kpc grid ``\texttt{explore\_cube\_density\_values\_025pc\_v2.fits}''. In
order to make these consistent to first order with the extinction maps
of \citet{2019ApJ...887...93G}, we apply the following corrections: i)
$A_{V} = A_{5500}/0.978$; ii) $A_{V}|_{A_{V} \geq 0.15} = A_{V}|_{A_{V}
\geq 0.15}+0.05$, and, iii) $A_{V}|_{A_{V} \leq 0.15} = (4/3)\times
A_{V}|_{A_{V} \leq 0.15}$, with the errors subsequently inflated to
$\sigma_{A_{V}}=0.31 \times A_{V}$.  Finally, for stars for which
neither extinction map provides measurements, we use the SFD maps
\citep{1998ApJ...500..525S}, with the value of $A_{V} = 2.742 E(B-V)$
suggested by the recalibration of \citet{2011ApJ...737..103S}.
\end{enumerate}
For each of these parameters, we require that the data satisfy quality
flags which we describe in the Appendix. We also inflate all photometric
errors by the arbitrary value of 0.01 mag in quadrature.

\subsection{Contributions to the Likelihood Function}

We construct our likelihood function with the aim of incorporating as
many reliable measurements as possible, but to do so without redundancy,
as many of the measurements probe nearly identical wavelength regimes. We
thus use the following inclusion criteria for our likelihood function:
\begin{enumerate} [itemsep=0.0cm]
\item We require at least one measurement of \{$NUV$, $u_{\rm{SM}}$,
$u_{\rm{SDSS}}$\}.
\item If measurements of both of \{$u_{\rm{SM}}$, $u_{\rm{SDSS}}$\}
are available, we use that of $u_{\rm{SDSS}}$.
\item If a measurement of $v_{\rm{SM}}$ is available, we use it.
\item We require a measurement of $G$, and we use it.
\item We use whichever has the most available measurements of
\{$G_{\rm{BP}}$,$G_{\rm{RP}}$ \}, \{$g_{\rm{SM}}$, $r_{\rm{SM}}$,
$i_{\rm{SM}}$, $z_{\rm{SM}}$\}, \{$g_{\rm{SSDSS}}$, $r_{\rm{SDSS}}$,
$i_{\rm{SDSS}}$, $z_{\rm{SDSS}}$\}, and \{$g_{\rm{PS}}$, $r_{\rm{PS}}$,
$i_{\rm{PS}}$, $z_{\rm{PS}}$\}, where ''PS" stands for Pan-STARRS. If
the number of available measurements are equal, we prioritize those of
Pan-STARRS, then SDSS, then Skymapper, then Gaia.
\item If a measurement of $y_{\rm{PS}}$ is available, we use it.
\item We require and use measurements for each of 2MASS \{$J$, $H$,
$K_{s}$ \} and WISE \{ $W_{1}$, $W_{2}$ \}.
\item If the extinction measurement is from either the PS1 maps or the
\citep{2022A&A...661A.147L,2022A&A...664A.174V} maps, we use it with
the associated measurement errors, where the latter is as described in
Section \ref{subsec:data}. If the extinction measurement is from the
SFD maps, it does not contribute to the likelihood.
\item We use the parallaxes and associated uncertainties from the Gaia DR3
catalog, with the zero-point corrections from \citet{2021A&A...649A...4L}.
\end{enumerate}

We keep track of all photometric measurements and model predictions
thereof, even those not contributing to the likelihood functions, for
purposes of subsequent heuristic comparisons.

\subsection{Contributions to the Priors}

We use the following priors for the exploration of the parameter space
for each star:

\begin{itemize}
    \item A flat prior in metallicity, over the interval $-2.0 \leq \rm{[Fe/H]} \leq + 0.50$.
    \item A flat prior in age in the interval $1 \leq \tau/\rm{Gyr} \leq 13.721$.
    \item A Chabrier prior in the mass of the star, restricted to the range $0.70 \leq m/m_{\odot} \leq 2.0$.
    \item A flat prior in extinction over the interval $A_{V} - 3\delta_{A_{V}} \leq A_{V} \leq A_{V} + 3\delta_{A_{V}}$, where $\delta_{A_{V}} = \sqrt{\sigma_{A_{V}}^2+0.01^2}$. Here, $\sigma_{A_{V}}$ is either the value given for the uncertainty in extinction from the PS1 or \citet{2022A&A...661A.147L}/\citet{2022A&A...664A.174V} maps, if those are used, or equal to 10\% of the extinction if we use the value from the SFD maps. If the value of $A_{V} - 3\delta_{A_{V}}$ is negative, the lower bound on the extinction is set to 0.
    \item A prior in distance that is proportional to the square of the distance (and thus uniform in volume), but restricted to the range $1/(\pi + 3\sigma_{\pi}) \leq d \leq 1/(\pi - 3\sigma_{\pi})$.
\end{itemize}

\begin{figure*}
\includegraphics[width=0.95\textwidth]{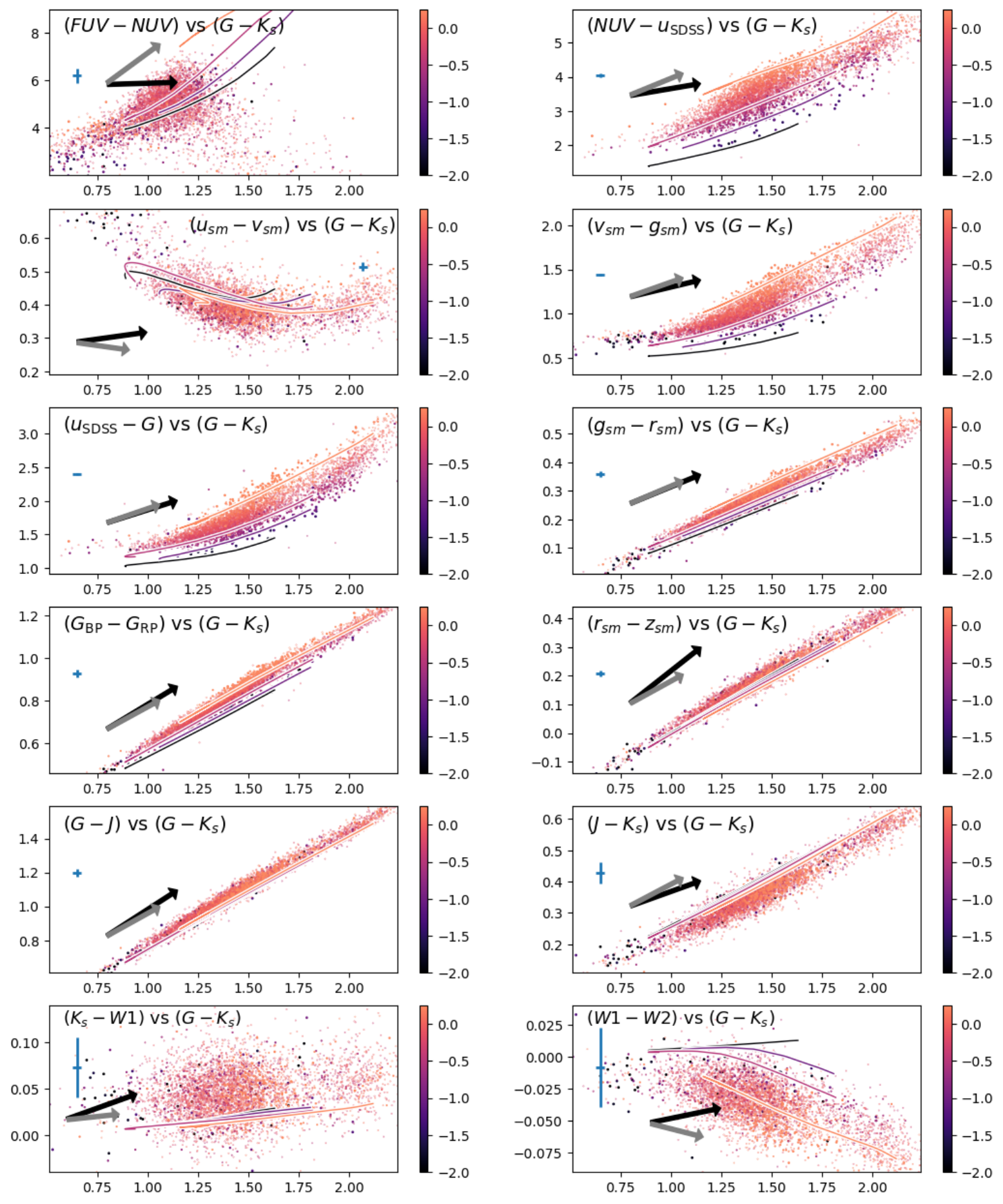}
\caption{We show various photometric scatters for stars on or near the
subgiant branch. The [Fe/H] measurements \citep{2021MNRAS.506..150B}
are color-coded, as are the predictions from the MIST isochrones. The
effects of increasing extinction by ${\Delta}A_{V}=0.50$ mag
and ${\Delta}T_{\rm{eff}}=500 \, K$ are approximated by the
black and grey arrows respectively, and representative photometric
uncertainties are shown by the cyan error bars. The association between
spectroscopically-inferred metallicities and color-color relations is
consistent with the theoretical prediction.}
\label{fig:GALAH}
\end{figure*}

\section{Methodology and Validation Thereof} \label{sec:Clusters}

We compute the posteriors for the parameters of each
star in our sample using the \texttt{isochrones} package
\citep{2015ascl.soft03010M}\footnote{\url{https://isochrones.readthedocs.io/en/latest/}},
which uses the PyMultinest implementation \citep{2014A&A...564A.125B}
of \texttt{MultiNest} \citep{feroz2008,feroz2009,feroz2019} to
compare astrometric, photometric, and spectroscopic observations
of stars to predictions from the MIST isochrones. The isochrones
package has now been widely used to estimate stellar parameters (e.g.
\citealt{2015ApJ...809...25M,2016Natur.533..509M,2016ApJS..224....2H,2022AJ....163..159R})
as has MultiNest in other contexts such as microlensing
(e.g. \citealt{2017A&A...604A.103P}), where it has been demonstrated to
be effective and efficient in the exploration of degenerate likelihood
spaces as well as those with multiple modes.

The isochrones package fully explores the MIST isochrones
within the bounds of the priors, by varying the distance, age,
initial mass, [Fe/H], and $A_{V}$ at which the isochrones are
evaluated. That is done by interpolating in the variable EEP (equivalent
evolutionary point), and using the bolometric correction tables from
MIST\footnote{\url{https://waps.cfa.harvard.edu/MIST/model\_grids.html\#bolometric}}.

We use seven different methods to validate and investigate our
methodology, which we discuss below.

\subsection{Validation of the Diagnostic Potential of Photometric
Measurements Along the Subgiant Branch } \label{subsec:GALAH}

In Figure \ref{fig:GALAH}, we plot the scatters of various
photometric color measurements of stars in the GALAH survey
\citep{2015MNRAS.449.2604D,2021MNRAS.506..150B}. These are selected to be
on or near the subgiant branch ($3.25 \leq \log{g} \leq 4.25$), to have
relatively reliable spectroscopic metallicity determinations ($-2.00 \leq
\rm{[Fe/H]} \leq +0.50$, $\sigma_{\rm{[Fe/H]}} \leq 0.20$), and to have
low extinctions ($A_{Ks} \leq 0.03$). We compare these to the predictions
from the MIST isochrones with ([Fe/H],$\tau/\rm{Gyr}$)$=(-2.0,10)$,
$(-1,10)$, $(-0.50,4)$, and $(+0.25,4)$ with $3.25 \leq \log{g} \leq
4.25$.

We draw several conclusions from this comparison. The first is that
measurements in $NUV$, $u$ and $v$ are particularly sensitive to
metallicity. The second is that the black arrow ($\delta A_{V} = 0.50$
mag) is usually nearly parallel and approximately 30\% longer than the
grey arrow  ($\delta T_{\rm{eff}} = 500$ K), and thus the uncertainties
between the temperature and reddening measurements will usually be
correlated. That is why constraints from extinction maps are arguably
necessary for our analysis. The third is that the data are slightly offset
from the predictions, by a few hundredths of a magnitude, in $(G-J)$,
$(J-K_{s})$, and $(K_{s}-W_{1})$.

The fourth, and most substantial, conclusion that we derive is that
there is an impressive match between the metallicity sensitivity of the
color-color relations predicted by the MIST isochrones, and that which
can be measured from the GALAH data. That need not have been the case,
as the latter are determined by high-resolution spectroscopy, whereas the
former are predicted by the product of model atmospheres with estimates of
the photometric transmission curves. The qualitative demonstration
of this consistency in Figure \ref{fig:GALAH} satisfies two requirements
for our methodology to be successful -- that broadband colors of stars
on the subgiant branch encode some information on metallicity, and that
the MIST isochrones reliably predict the trends.

We note that it has long been known that photometry can
be incredibly constraining in the inference of stellar parameters (e.g.
\citealt{1953ApJ...117..313J,1966ARA&A...4..433S,2002MNRAS.337..151H,2011A&A...530A.138C}).
Where this study differs is our emphasis on the subgiant branch to
infer ages as well, and the combination of the most widely available
ultraviolet through infrared photometry and astrometry for these stars.

\subsection{Validation of Astrophotometrically-inferred Metallicities
via Comparisons to Spectroscopic Surveys.} \label{subsec:Spectroscopy}

\begin{figure*}
\centering
\includegraphics[width=0.9\textwidth]{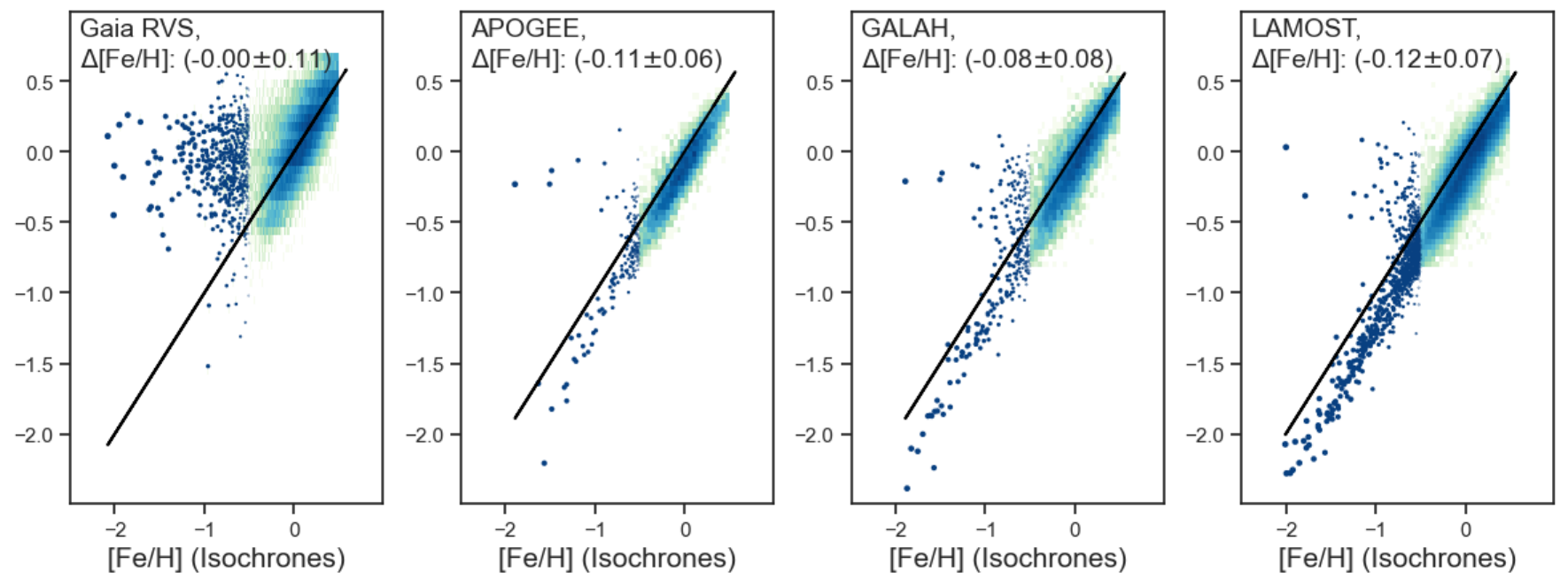}
\caption{The comparison between  our astrophotometric metallicity
determines and those from four major spectroscopic surveys. Legends
show the median difference in [Fe/H], where a positive value denotes
a higher metallicity for the spectroscopic determination, and the
median absolute deviation between the metallicity determinations.
The black lines denote the lines of equality between the metallicity
determinations. This comparison to data from four surveys demonstrated
that our astrophotometric estimates of metallicity are largely consistent
with the spectroscopically-inferred values.}
\label{fig:MetComparisons}
\end{figure*}

We compare our astrophotometric metallicity determinations
to those from four major spectroscopic surveys in Figure
\ref{fig:MetComparisons}. These four surveys are the Gaia High-Resolution
Spectrograph Survey \citep{2023AnA...674A...1G}, the APOGEE survey
(DR17, \citealt{2017AJ....154...94M}), the GALAH survey (DR3,
\citealt{2021MNRAS.506..150B}), and the LAMOST survey (LRS DR7,
\citealt{2015MNRAS.448..822X}), where we describe our inclusion criteria
for these data in our appendix.

These comparisons validate the assumption that we can in fact reliably
infer metallicities over a broad metallicity range, $-2.0 \leq \rm{[Fe/H]}
+0.50$. Our metallicity inferences are approximately 0.10 dex lower than
the spectroscopic values, with a median absolute deviation between the
two values of approximately 0.10 dex.

We evaluate the comparison with the GALAH data in greater
detail. For those stars, the median offset on the [Fe/H] determinations
is 0.08 dex, with a median absolute deviation of 0.08 dex. That median
offset is slightly larger than the median reported measurement error
in our astrophotometrically-derived metallicities (0.06 dex), or the
spectroscopically-derived metallicities from GALAH (0.07 dex).

In each of the four panels, we see cloud of points with low
astrophotometric metallicities but with high spectroscopic metallicities
of [Fe/H] $\geq -0.50$. For the APOGEE, GALAH, and LAMOST sample we
consider it likely that this discrepancy is due to issues with our
photometric analysis rather than with the spectroscopic analysis. The
reasoning for this is presented in Section \ref{subsec:MPannex}, where
we are able to run tests on a larger number of such stars.

\subsection{Validation of the Precision and Accuracy of Ages and
Metallicities With Well-studied Open Clusters} \label{subsec:OpenClusters}

The open clusters M67 (NGC 2682) and NGC 188 are well studied, and thus
provide an independent means to validate our methodology to infer ages and
metallicities. We can evaluate if our parameter inferences for subgiant
stars within a cluster are consistent with one another as a proxy for
precision, and if they are consistent with other literature values as
a proxy for accuracy.

\subsubsection{Empirical Population Parameters for the Validation
Clusters} \label{subsec:Clusters}

All literature references for cluster parameters that either inform our
choice of priors or the evaluation of our results are listed in Table
\ref{Table:Clusters}.

For the distances to the clusters, we assume the inverse of the
parallax values that have been vetted by the Gaia collaboration
\citep{2018AnA...616A..10G}. Both parallaxes are precisely measured
($\pi/\sigma_{\pi} \geq 100$), and we shift the parallaxes by a
zero-point offset of $\delta_{\pi}=0.054$ mas, such that the parallaxes
are increased \citep{2019MNRAS.487.3568S}. In contrast, if we used the
parallax measurements for each individual stars, we would have a median
precision of $\pi/\sigma_{\pi}  \approx 70$ for M67 and $\pi/\sigma_{\pi}
\approx 30$ for NGC 188, and thus the parameter inferences would be
degraded relative to most of our sample.

For the extinction to these clusters, the Pan-STARRS reddening maps
\citep{2019ApJ...887...93G} would be the ideal choice for reddening
estimates, as using these values would provide the most consistency with
the rest of our work. However, we have found that these estimates are
unphysically noisy toward clusters, varying by a factor of several among
the cluster stars. That may be because of confusion affecting some of
the photometry used to construct those maps, or because a pile-up of
stars at a specific distance radically violates the smooth priors on
the distribution of stars in the Milky Way assumed in the construction
of those maps (Gregory Green, private communication). We instead use the
inferences $A_{V,\rm{M67}}=0.120 \pm 0.012$ and $A_{V,\rm{NGC\,188}}=0.260
\pm 0.026$ as part of our likelihood, with the priors on the extinction
modified as they are for stars in the main sample.

Literature and age estimates of the metallicities of these clusters
vary. For these parameters, we assume the same priors as we do for the
stars in the main sample, a flat prior in metallicity over the range
$-2.0 \leq \rm{[Fe/H]} \leq +0.50$ and a flat prior in $\log{\rm{age}}$
over the range $1.0 \leq \tau/\rm{Gyr} \leq 13.721$. As elsewhere, we
assume a Chabrier prior in initial stellar mass over the range $0.70
\leq M/M_{\odot} \leq 2.0$

The inclusion criteria for cluster stars in terms of photometry and
indicators of (non)-variability are the same as they are for stars in
the main sample, yielding 20 stars in M67 and 34 stars in NGC 188.

\begin{deluxetable*}{  llllll }
\tablecaption{Literature references as well as the estimates in this
work for the population parameters of the clusters. Where references
report reddening in terms of $E(B-V)$, we multiply their values by
$A_{V}/E(B-V)=3.1$. The inclusion criteria for cluster stars in terms
of photometry and indicators of (non)-variability are the same as they
are for stars in the main sample, yielding 20 stars in M67 and 34 stars
in NGC 188. \label{Table:Clusters}}
\tablehead{
\colhead{Population} & \colhead{parallax/mas} & \colhead{[Fe/H]} & \colhead{$\log \rm{(Age)}$} &   \colhead{$A_{V}$} & \colhead{Reference and Comments} }
\startdata
M67  & 1.1865 $\pm$ 0.0011 & $\cdots$ & $\cdots$ & $\cdots$ & \citet{2018AnA...616A..10G,2019MNRAS.487.3568S} \\
M67 & $\cdots$ & $+0.01$ & $\cdots$ & $\cdots$ & \citet{2014AnA...561A..93H} \\
M67  & $\cdots$ & $+0.00$ & $\cdots$ & $\cdots$ & \citet{2019AnA...627A.117L} \\
M67  & $\cdots$ & $+0.03$ & $\cdots$ & $\cdots$ & \citet{2019MNRAS.490.1821C} \\
M67 & $\cdots$ & $+0.01$ & $\cdots$ & $\cdots$ & \citet{2020AJ....159..199D} \\
M67  & $\cdots$ & $+0.00$ & $\cdots$ & $\cdots$ & \citet{2021MNRAS.503.3279S} \\
M67  & $\cdots$ & $\cdots$ & 9.32 & $\cdots$ & \citet{2013AnA...558A..53K} \\
M67  & $\cdots$ & $\cdots$ & 9.63 & $\cdots$ & \citet{2020AnA...640A...1C} \\
M67  & $\cdots$ & $\cdots$ & 9.54 & $\cdots$ & \citet{2016ApJ...832..133S} \\
M67  & $\cdots$ & $\cdots$ & 9.63 & $\cdots$ & \citet{2019AnA...627A.117L} \\
M67  & $\cdots$ & $\cdots$ & 9.57 & $\cdots$ & \citet{2021AJ....161...59S} \\
M67  & $\cdots$ & $\cdots$ & $\cdots$ & 0.124 & \citet{1999AJ....118.2894S} \\
M67  & $\cdots$ & $\cdots$ & $\cdots$ & 0.186 & \citet{2005AnA...438.1163K} \\
M67  & $\cdots$ & $\cdots$ & $\cdots$ & 0.186 & \citet{2010MNRAS.403.1491P} \\
M67  & $\cdots$ & $\cdots$ & $\cdots$ & 0.27 & \citet{2009AJ....137.5086M} \\
M67  & $\cdots$ & $\cdots$ & $\cdots$ & 0.127 & \citet{2007AJ....133..370T} \\
M67  & $\cdots$ & $\cdots$ & $\cdots$ & 0.109 & \citet{2017EPJWC.16005005V} \\
M67 & $\cdots$ & $\cdots$ & $\cdots$ & 0.093 & \citet{2022JAsGe..11..166H} \\
M67 & $\cdots$ & $\cdots$ & $\cdots$ & 0.248 & \citet{2007ApJ...666L.105V} \\
NGC 188 & 0.5593 $\pm$ 0.0011  & $\cdots$ & $\cdots$ & $\cdots$ &  \citet{2018AnA...616A..10G,2019MNRAS.487.3568S}  \\
NGC 188 & $\cdots$ & $+0.11$ & $\cdots$ & $\cdots$ & \citet{2014AnA...561A..93H} \\
NGC 188 & $\cdots$ & $+0.03$ & $\cdots$ & $\cdots$ & \citet{2019MNRAS.490.1821C} \\
NGC 188 & $\cdots$ & $+0.09$ & $\cdots$ & $\cdots$ & \citet{2020AJ....159..199D} \\
NGC 188 & $\cdots$ & $+0.09$ & $\cdots$ & $\cdots$ & \citet{2021MNRAS.503.3279S} \\
NGC 188 & $\cdots$ & $\cdots$ & 9.65 & $\cdots$ & \citet{2013AnA...558A..53K} \\
NGC 188 & $\cdots$ & $\cdots$ & 9.85 & $\cdots$ & \citet{2020AnA...640A...1C} \\
NGC 188 & $\cdots$ & $\cdots$ & 9.84 & $\cdots$ & \citet{2016AJ....152..129C} \\
NGC 188 & $\cdots$ & $\cdots$ & $\cdots$ & 0.279 & \citet{1999AJ....118.2894S} \\
NGC 188 & $\cdots$ & $\cdots$ & $\cdots$ & 0.248 & \citet{2010MNRAS.403.1491P} \\ \hline
M67  & 1.1897 $\pm$ 0.0007 & 0.04 $\pm$ 0.08 & 9.58 $\pm$ 0.04 & 0.12 $\pm$ 0.01 & Derived means and weighted standard deviations \\
M67  & 1.1865 $\pm$ 0.0000 & 0.03 $\pm$ 0.05 & 9.60 $\pm$ 0.01 & 0.12 $\pm$ 0.00 & Medians and median absolute deviations \\
M67 \citep{2022Natur.603..599X}  &  $\cdots$ & $-0.11 \pm  0.05$ & 9.63 $\pm$ 0.03 & $\cdots$  & Derived means and  standard deviations \\
M67 \citep{2022Natur.603..599X}  &  $\cdots$ & $-0.11 \pm  0.02$  & 9.61 $\pm$ 0.02 & $\cdots$  & Medians and median absolute deviations \\
NGC 188 & 0.5593 $\pm$ 0.0000 & 0.11 $\pm$ 0.18 & 9.79 $\pm$ 0.08 & 0.26 $\pm$ 0.03 & Derived means and standard deviations \\
NGC 188 & 0.5593 $\pm$ 0.0001 & 0.13 $\pm$ 0.06 & 9.80 $\pm$ 0.03 & 0.26 $\pm$ 0.06 & Medians and median absolute deviations \\
\enddata
\end{deluxetable*}

\subsubsection{Ages of Cluster Subgiants: Trends and Results}
\label{subsec:ClusterAges}

\begin{figure*}
\centering
\includegraphics[width=0.90\textwidth]{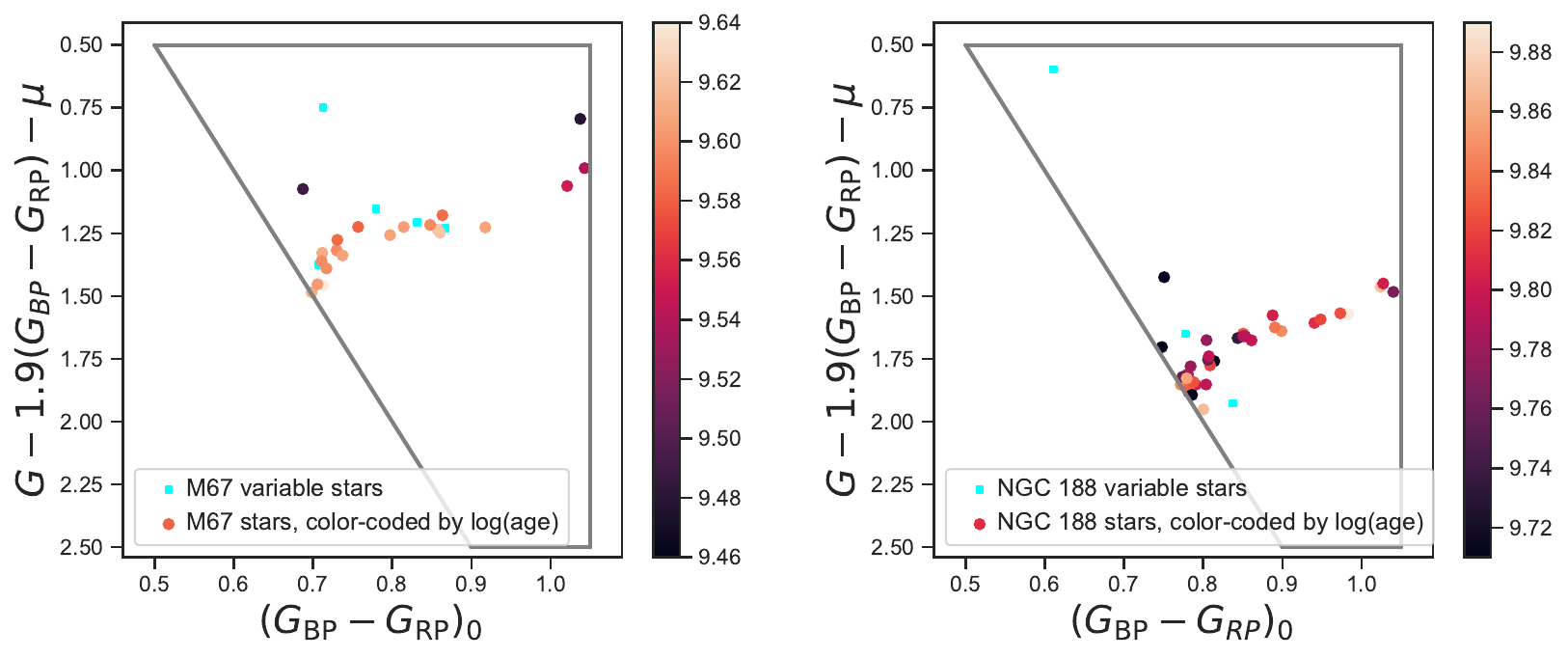}
\caption{Color-magnitude diagrams of stars in M67 (LEFT) and NGC 188
(RIGHT) color-coded by derived age, with the stars identified as variable
or otherwise less reliable shown as cyan squares. Our analysis finds
very little dispersion in derived ages among the subgiant branch stars,
but there is some contamination from stars that are likely be blue
stragglers.}
\label{fig:Clusters}
\end{figure*}

Our results for stars associated with M67 and NGC 188 are summarized
at the bottom of Table \ref{Table:Clusters}, and the dependence of
age on location in the color-magnitude diagram is  shown in Figure
\ref{fig:Clusters}.

The derived metallicities and ages are both consistent with literature values, and consistent with one another. For M67, the median absolute deviation and 1-$\sigma$ precision in ages and metallicities are 3\%, 9\% and 0.05, 0.08 dex respectively. For NGC 188, the median absolute deviation and 1-$\sigma$ precision in ages and metallicities are 6\%, 19\% and 0.06, 0.18 dex respectively.


For both clusters, the median absolute deviations in the ages are
substantially smaller ($\approx 70\%$) than the weighted standard
deviations, where we would only expect them to be $\approx 33\%$ smaller
for normally distributed errors. That is because the distributions of
derived parameters have a high kurtosis as the ends of the distributions
are dominated by outliers. The nature of those outliers are easy to
identify in Figure \ref{fig:Clusters} -- our variability criteria can flag
many, but not all, of the stars that are binaries, products of binary
evolution, et cetera. There may also be some field star contamination,
even after selecting for stars using proper motions.

The subgiant stars in NGC 188 are $\approx 3 \times$ less likely to be
identified as variables than those in M67 --and that is not surprising,
as some of of our variability criteria will be less effective for stars
at the greater distance of NGC 188 (1788 versus 843 pc). The brightest
unflagged bright subgiant stars in M67 and NGC 188, that may very well be
blue stragglers, respectively have derived ages of $\log(\rm{age})=9.48$
and $\log(\rm{age})=9.64$. These stars add to the dispersion in derived
parameters of these clusters in a manner that will be less likely for
nearby field stars.

A key limitation of this comparison, as pointed out by the
anonymous referee, is that stellar isochrones and the stellar physics
that go into them have been evaluated and even calibrated by comparison
to clusters, and thus these comparisons are not necessarily
independent. In this particular case, \citet{2016ApJ...823..102C} actually fit for a key parameter of stellar models, the convective overshoot for stellar cores, by comparing the shape of the main-sequence turnoff in M67 to the predictions. They derived  $f_{\rm{ov,core}}=0.016$.

\subsection{Comparison to the results of \cite{2022Natur.603..599X}
}\label{sec:Comparison}

The work of \cite{2022Natur.603..599X} is arguably the most similar
comparison that we have to our own investigation. They too, sought to
use the tremendous diagnostic power of Gaia astrometry to estimate ages
of subgiant branch stars, and by extension, of the solar neighborhood
as a whole. There are several differences in our methodologies, most
particularly:
\begin{itemize}
\item
For the most effective constraint on metallicity, our investigation
requires an ultraviolet flux measurement, whereas that of
\cite{2022Natur.603..599X} required a LAMOST spectra and subsequently
derived stellar parameters.
\item
Our method estimates a mean total metallicity as ``[Fe/H]" whereas that
of \cite{2022Natur.603..599X} splits the metallicity into [Fe/H] and
[$\alpha$/Fe] parameters. The latter approach comes with the ``cost" of an
extra free parameter, but it is one which is physically well-motivated
and constrained by the LAMOST spectra. The impact of accounting for
[$\alpha$/Fe] is shown in the right panel of Supplementary Figure 4 of
\cite{2022Natur.603..599X}: Isochrones that underestimate [$\alpha$/Fe]
by 0.20 dex will overestimate ages by approximately 1.5 Gyr in the mean,
but that is in the case where [Fe/H] is fixed to the spectroscopic
value. In our case, we infer an ``effective" [Fe/H] value, which will
be shifted for $\alpha$-enhanced stars.
\item
Both methods derive $A_{V}$ from available extinction maps, predominantly
that of \citet{2019ApJ...887...93G} in this investigation and entirely
in that of \citet{2022Natur.603..599X}. However, our method assumes  a
constant total-to-selective extinction ratio $R_{V}=A_{V}/E(B-V)=3.1$,
whereas that of \cite{2022Natur.603..599X} varies $R_{V}$ for every star
as an additional free parameter.
\item
The two investigations have different variability exclusion criteria,
which are too numerous to fully describe here.
\end{itemize}

There are 31,876 stars in common between our two samples,
including 29,965 that satisfy our sample's variability and blending
criteria. Among those, our derived [Fe/H] values are systematically
higher, with ${\Delta}\rm{[Fe/H]} \sim 0.19 \pm 0.10$, and our derived
ages are systematically lower, with the age ratio distributed as
$\tau_{\rm{This\,Work}}/\tau_{\rm{Xiang}} \sim 0.94 \pm 0.13$.

A diagnostically powerful point of comparison is that of the open
cluster M67, for which there are numerous available  literature
benchmark measurements. Our sample includes 20 M67 members, that of
\citet{2022Natur.603..599X} includes 7 member stars, and there are four
stars in common between our two samples. Of the three that do not make
it into our sample, two are excluded because they are bluer than our
subgiant selection cutoff, and one is excluded due to the absence of an
ultraviolet photometric measurement.

Regardless of whether or not we select the 4 stars in common or
the 7 total M67 stars in their sample, \citet{2022Natur.603..599X}
infer a slightly higher mean age for M67, and a slightly lower mean
metallicity. The respective ratio (95\%) and offset (0.14 dex) are
effectively identical to that for for our samples as a whole.

We conclude our two samples are consistent, showing that ultraviolet
photometry and large-survey spectroscopy currently have comparable
diagnostic power for the inferences of ages and metallicities of subgiant
branch stars.

\subsection{Validations of the Assumed Extinction Maps}

In this investigation we assume the
extinction maps of  \citet{2019ApJ...887...93G} and
\citet{2022A&A...661A.147L}/\citet{2022A&A...664A.174V}. When extinction
estimates from both maps are available, we prioritize those from
the former. Where only measurements from the latter are available,
we transform them using the prescription in item \#8 of the list in
Section \ref{subsec:data}, which we derive here.

Here, we seek to assess the validity of those assumptions as follows. We
use a subsample of stars for which we have an extinction estimate from
both reddening maps, for which GALEX $NUV$ is measured to a precision
of better than 0.10 mag, and for which there is also an ultraviolet
measurement from either Skymapper or SDSS. That yields a sample of
18,176 stars.

For those stars, the combination of both $NUV$ and $u$ photometry
should enable us to reliably estimate extinction from the photometry
and astrometry alone. We thus remove the extinction measurements from
the likelihood, and relax the prior to being a flat prior in the range
$A_{V} \in [0, A_{\rm{V,Green}}+A_{\rm{V,Vergely}}+0.05]$.

For this sample, we find that the extinction maps of
\citet{2019ApJ...887...93G} fare better than those of
\citet{2022A&A...664A.174V} in the mean offsets, with offsets of
$A_{\rm{V,Maps}} - A_{\rm{V,Inferred}}$ of $0.035 \pm 0.092$ versus $0.066
\pm 0.100$ respectively. Here the errors denote 1-$\sigma$ dispersions.

That may however be an artifact of our sample -- we notice, for example,
that the maps of \citet{2022A&A...664A.174V} fare better than of
\citet{2019ApJ...887...93G} for lower values of inferred extinction. If
we restrict the comparison to $A_{\rm{V,Inferred}} \leq 0.10$, the mean
offsets are $0.024 \pm 0.058$ and $0.038 \pm 0.079$ mag respectively.

\begin{figure*}
\centering
\includegraphics[width=0.90\textwidth]{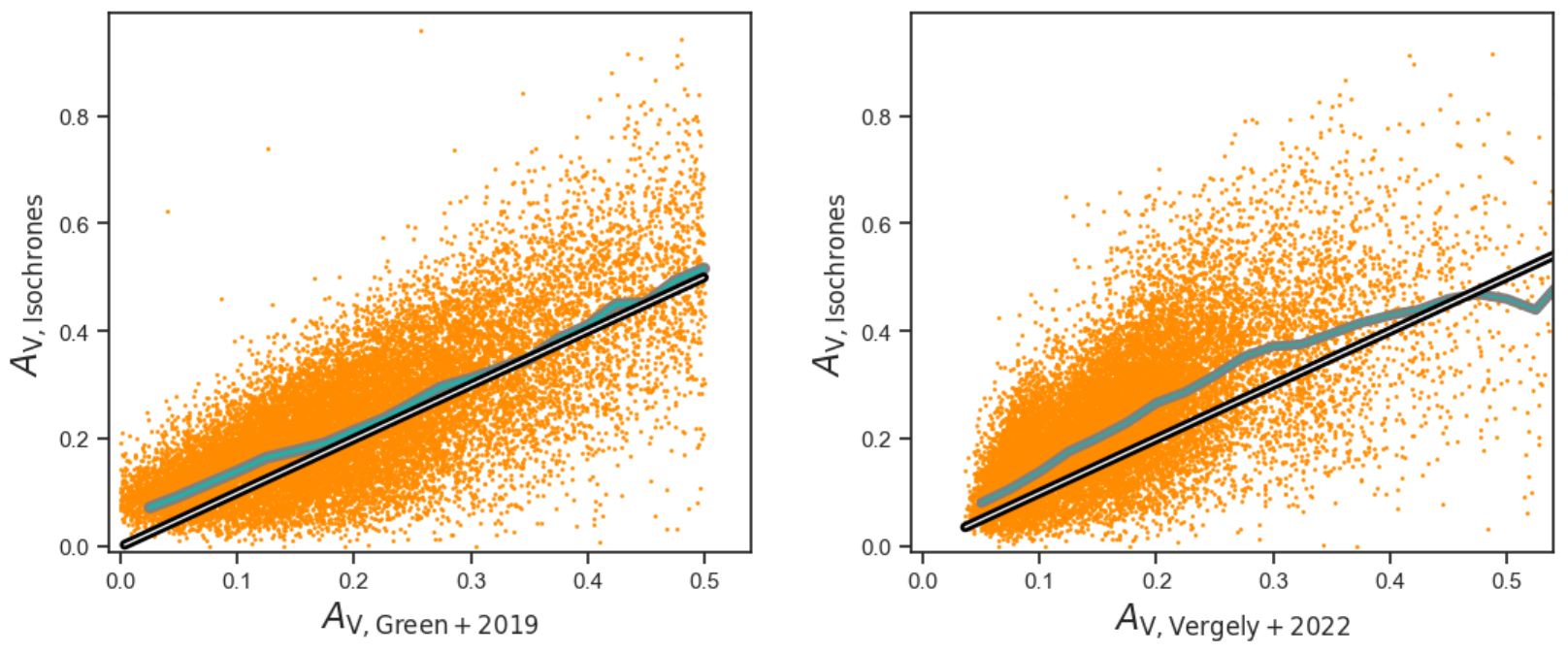}
\caption{Comparison of our derived extinction values relative
to those of the \citet{2019ApJ...887...93G} maps (LEFT) and the
\citet{2022A&A...664A.174V} maps (RIGHT). The black and white line denotes
equality, and the blue line denotes the median trends.  Both maps do
well in the mean, but the predictions of \citet{2019ApJ...887...93G}
fare better for $A_{V} \gtrapprox 0.10$, which comprise the majority of
our primary sample.}
\label{fig:ReddeningMaps}
\end{figure*}

\subsection{Validation of the Assumed Bolometric Coefficients and
Extinction Curve}

The uncertainties of the assumed bolometric corrections and interstellar
extinction curves are among the most obvious contributors to the
uncertainty in photometric investigations such as this one.

In order to estimate the magnitude of this uncertainty, we construct
a subsample of 5,558 stars which have: i) a GALEX NUV magnitude with
measurement precision better than 0.10 mag; ii) a $u$-band measurement;
iii) a precision in  extinction from \citet{2019ApJ...887...93G} that
is better than 10\%; and, iv) a metallicity measurement from LAMOST
with [Fe/H]$\leq +0.35$. For the latter, we then assume a metallicity
measurement of [Fe/H]=[Fe/H]$_{\rm{LAMOST}}+0.12$ in our likelihood
(see Figure \ref{fig:MetComparisons}), and we inflate all photometric
uncertainties by 0.05 in quadrature, rather than the 0.01 mag used
elsewhere in this text.

\begin{figure}
\centering
\includegraphics[width=0.49\textwidth]{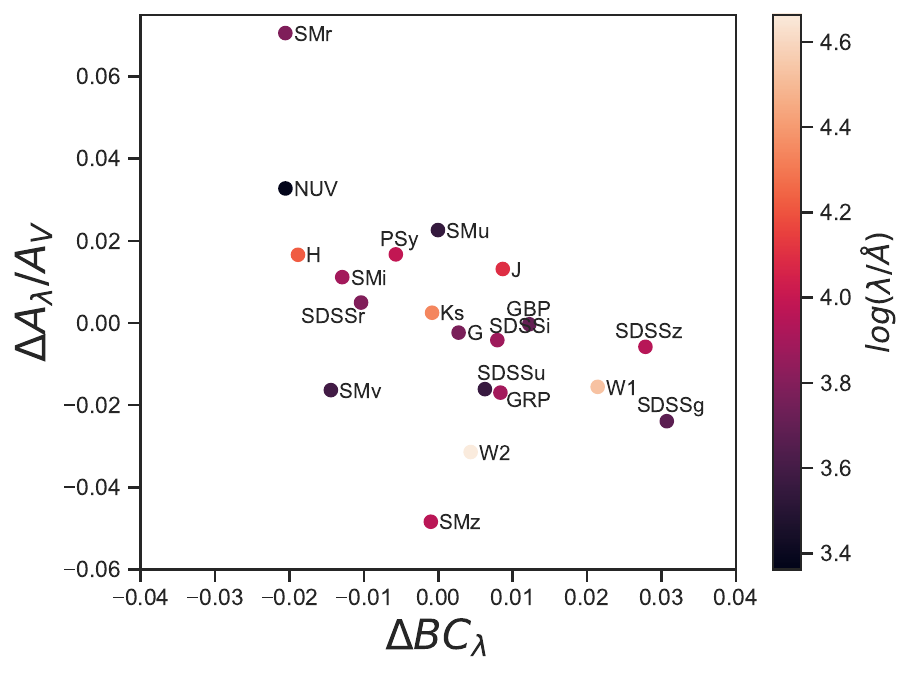}
\caption{The scatter of derived errors in the mean extinction coefficients
as a function of the derived mean errors in the bolometric corrections,
color-coded by filter effective wavelength. The fact that the two
variables are anti-correlated is indicative of  undiagnosed systematic
errors, such as errors in the assumed filter transmission curves.}
\label{fig:BolometricCorrections}
\end{figure}

Then, for each bandpass X, we plot the scatter of $X_{\rm{Predicted}} -
X_{\rm{Measured}}$ as a function of $A_{V}$, and compute the least-squares
linear fit. We then show the slopes as a function of the intercepts for
each bandpass in Figure \ref{fig:BolometricCorrections}.

In principle, this should be a scatter of the errors in the extinction
coefficients $\Delta A_{X}/A_{V}$ as a function of the errors in the
bolometric corrections $\Delta BC_{X}$ for each bandpass X, but in
practice we see that this is unlikely to be the case. The two variables
are anti-correlated, with a Spearman coefficient of $\rho=-0.53$ with
p-value of $p=0.02$. We consider this anti-correlation to be indicative
of small, undiagnosed systematic errors.

Another way to discern the presence of such errors is that the
coefficients for bandpasses of similar effective wavelengths, denoted
by similar color-coding in Figure \ref{fig:BolometricCorrections}, do
not yield the same derived errors in the extinction coefficients. For
example there is a 0.04 mag offset in $\Delta A_{X}/A_{V}$ between
$u_{\rm{SDSS}}$ and $u_{\rm{SM}}$, and a 0.07 mag offset in $\Delta
A_{X}/A_{V}$ between $r_{\rm{SDSS}}$ and $r_{\rm{SM}}$, when those
offsets should be close to 0 on physical grounds. Separately, it is known
that there is an uncertainty of 1-5\% in the zero-points of the GALEX
filters \citep{2010ApJ...725.1215S}, which should be reduced for upcoming
ultraviolet observatories such as ULTRASAT \citep{2023arXiv230414482S}
and UVEX \citep{2021arXiv211115608K}.

Regardless of these issues, it is at least the case that the offsets
are small. The median of the absolute values of the offsets in the
bolometric corrections and the extinction coefficients are given
by $\Delta BC_{X}=0.010$ and $\Delta A_{X}/A_{V}=0.016$. The latter's
impact is further reduced due to the fact that the median extinction of
our primary sample is $A_{V} \approx 0.15$ mag.

At this time, we cannot reliably discern if these offsets are due to
the photometric data reduction or in the predictions of the stellar
models, but this may become feasible as more data become available,
in particular data from different observatories measuring photometry
in nominally similar bandpasses. Given these uncertainties, we choose
not to investigate the effects of possible sightline-dependent
variations in the interstellar extinction curve, which can bias
the derived parameters of stars if not adequately accounted for
\citep{2005ApJ...632..227R,2012A&A...542A..74G,2015MNRAS.449.1171N,2021ApJ...910..121N,2024arXiv240414626A,2024arXiv240503068B}.

\subsection{Sensitivity of the Derived Parameters to Input Data and
Priors} \label{subsec:Sensitivity}

Here, we evaluate the sensitivity of our derived stellar parameters as a
function of the input data and priors. For the tests following the first
one, we use a randomly-selected sample of approximately 5,000 stars
selected to have $\pi/\sigma_{\pi} \geq 100$ and $\sigma_{A_{V}} \leq
0.05$. We list the effects of these changes in approximately decreasing
order of the size of their effects.

\subsubsection{The Effect of Removing Ultraviolet Measurements from
the Likelihood}

We evaluate the effect of removing the measurements of $NUV$,
$u_{\rm{SM}}$,  $v_{\rm{SM}}$ and $u_{\rm{SDSS}}$ from the likelihood. The
mean offset in metallicity, where we report the catalog value subtracted
from the adjusted value, is $\Delta \rm{[Fe/H]} = -0.02 \pm 0.17$,
and similarly, $\Delta \log(\rm{Age}) = -0.00 \pm 0.06$. The Pearson
correlation between these two offsets is $\rho=-0.81$.

Thus, in the mean, removing ultraviolet measurements negligibly biases the
results, but it does yield a very large and very correlated statistical
error for derived ages and metallicities.

\subsubsection{The Effect of Removing Extinction Priors from the
Likelihood and Adjusting Them in the Prior}

We evaluate the effect of removing the available measurements of
extinction from both the likelihood and priors, and replacing them with
a flat prior in $A_{V}$ over the range $[0,1]$.

The mean offset in metallicity, where we report the catalog value
subtracted from the adjusted value, is $\Delta \rm{[Fe/H]} = -0.07
\pm 0.08$, and similarly, $\Delta \log(\rm{Age}) = 0.08 \pm 0.11$. The
Pearson correlation between these two offsets is $\rho=-0.72$.

\subsubsection{The Effect of Removing Both 2MASS and WISE Data from
the Likelihood}

We evaluate the effect of removing the measurements of $J$, $H$, $K_{s}$,
$W_{1}$, and $W_{2}$ from the likelihood.

The mean offset in metallicity, where we report the catalog value
subtracted from the adjusted value, is $\Delta \rm{[Fe/H]} = 0.02 \pm
0.08$, and similarly, $\Delta \log(\rm{Age}) = -0.02 \pm 0.06$. The
Pearson correlation between these two offsets is $\rho=-0.70$.

\subsubsection{The Effect of Removing WISE Data from the Likelihood}

We evaluate the effect of removing the measurements of $W_{1}$ and $W_{2}$
from the likelihood.

The mean offset in metallicity, where we report the catalog value
subtracted from the adjusted value, is $\Delta \rm{[Fe/H]} = -0.01
\pm 0.02$, and similarly, $\Delta \log(\rm{Age}) = 0.01 \pm 0.02$. The
Pearson correlation between these two offsets is $\rho=-0.46$.

\subsubsection{The Effect of Adjusting the Prior on Metallicity}

We evaluate the effect of changing our metallicity prior, which is
a flat prior in metallicity, over the internal $-2.0 \leq \rm{[Fe/H]}
\leq + 0.50$, to the default prior  suggested by isochrones. The default
prior has a dominant peak near [Fe/H]$=0$, with a long tail to lower
metallicities\footnote{The default priors for the isochrones package
are described more fully at \urlA}.

The mean offset in metallicity, where we report the catalog value
subtracted from the adjusted value, is $\Delta \rm{[Fe/H]} = 0.01 \pm
0.02$, and similarly, $\Delta \log(\rm{Age}) = 0.00 \pm 0.02$. The Pearson
correlation between these two offsets is $\rho=-0.65$. Approximately
5\% of the sample stars have changes in either of [Fe/H] or log(age)
exceeding 0.05 dex, and approximately 1\% of the sample stars have changes
exceeding 0.10 dex. These shifts in derived parameters are represented by
the errors: Whereas the stars whose derived values of $\log(\rm{Age})$
shifted by less than 0.05 had mean errors in their ages of 9.3\%, those
for which $\log(\rm{Age})$ shifted by more than 0.05 had mean errors in
their ages of 28.4\%.

\subsubsection{The Effect of Increasing or Decreasing the Number of Live
Points Used in the Nested Sampling}

We evaluate the effect of increasing the number of live points used by
MultiNest for the fitting, from 1,000 to 3,000 live points, and then
from 1,000 to 300 live points.

When we increase the number of live points to 3,000, the shift
in the mean offset in metallicity, where we report the catalog value
subtracted from the adjusted value, is $\Delta \rm{[Fe/H]} = 0.00 \pm
0.00$, and similarly, $\Delta \log(\rm{Age}) = 0.00 \pm 0.01$. The largest
change in the derived value of [Fe/H] is 0.03 dex. For the derived values
of $\log(\rm{Age})$, some 0.2\% of the stars have parameter shifts of more than
0.05 dex, with the largest recorded shift being 0.08 dex.

Similar results are obtained when decrease the number of live
points to 300, the shift in the mean offset in metallicity, where we
report the catalog value subtracted from the adjusted value, is $\Delta
\rm{[Fe/H]} = 0.00 \pm 0.00$, and similarly, $\Delta \log(\rm{Age}) =
0.00 \pm 0.01$. The largest change in the derived value of [Fe/H] is
0.04 dex. For the derived values of $\log(\rm{Age})$, some 0.6\% of the stars
have parameter shifts of more than 0.05 dex, with the largest recorded
shift being 0.11 dex.

\section{Main Results} \label{sec:Results}

\subsection{Availability of the Derived Parameters and Uncertainties Thereof}

The table of derived stellar parameters (16th, 50th, and 84th
percentiles of the posteriors) of our program stars can be found in
Table \ref{Table:FullDataPrime} of the online edition of this paper.

The input tables for the 401,819 stars in the primary
sample and the metal-poor annex, including the cross-matches
to other surveys, the python Jupyter Notebook used to evaluate
them, associated ReadMe file, and links to the table of derived
parameters can be found in a github repository of this paper's first
author\footnote{\url{https://github.com/DavidMoiseNataf/Subgiants}}.

The data that support the findings of this study are
openly available in the Johns Hopkins Research Data Repository at
\url{https://doi.org/10.7281/T1/CGBG4F}. That includes the above-mentioned
data, as well as the full probabilistic posteriors for the derived
stellar parameters (e.g. [Fe/H], $A_{V}$, etc) of each star.

\subsection{Primary Sample: Ages and Metallicities}

\begin{figure*}
\epsscale{1.10}
\plottwo{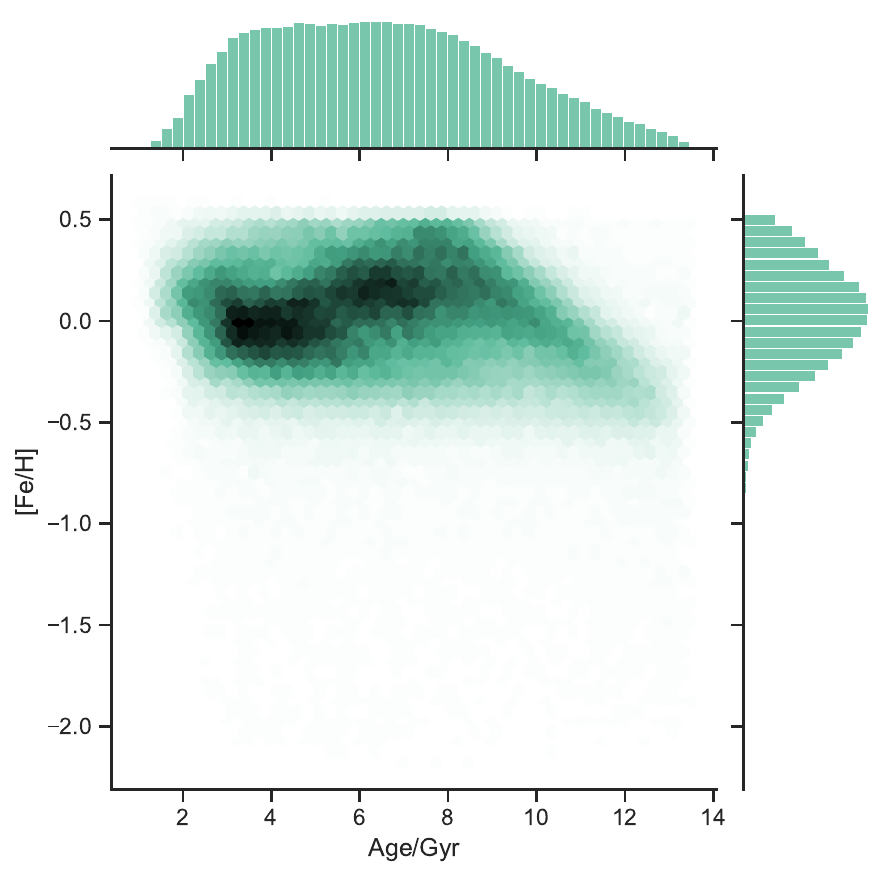}{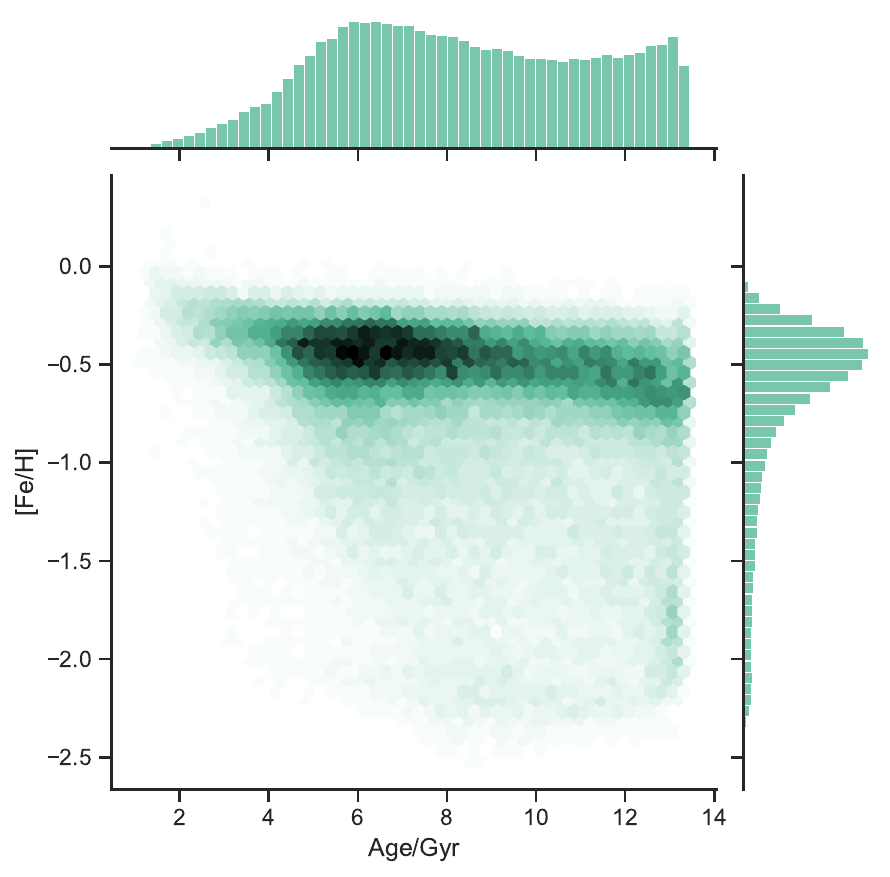}
\caption{LEFT: The distribution in age-metallicity space for 242,489
stars from our primary sample which are not associated with open clusters,
and which satisfy our other variability and blending inclusion criteria.
RIGHT: The distribution in age-metallicity space for 96,247 stars from
our metal-poor annex which are not associated with open clusters, and
which satisfy our other variability and blending inclusion criteria. The
age-metallicity distribution functions show considerable structure that
would not be discernible in either of the marginal distributions. }
\label{fig:AgeMetDistributions}
\end{figure*}

We show, in the right panel of Figure \ref{fig:AgeMetDistributions},
the distribution in age-metallicity space for 242,489 stars from our
primary sample which are not associated with open clusters, and which
satisfy our other inclusion criteria.

We observe three age-metallicity components in our sample. The first
is an excess concentration centered at  $(\tau/\rm{Gyr}, \rm{[Fe/H]})
\approx (4,0)$, corresponding to the age and metallicity of the Sun. The
second excess concentration is centered at $(\tau/\rm{Gyr}, \rm{[Fe/H]})
\approx (7,+0.15)$, and the third is a streak of stars distributed along
a narrow band in age-metallicity space, from $(\tau/\rm{Gyr}, \rm{[Fe/H]})
\approx (8,+0.15)$ to $(13,-0.50)$. Not one of these three features would
be discernible if we were to look at either of the marginal distributions
of age and [Fe/H] distributions, but they are easily discernible in the
joint distribution of age and metallicity.

We have verified that the presence of these three components is robust
to the following methodological changes : requiring an uncertainty
in $A_{V}$ of less than 0.05 mag; requiring a low inferred value for
the extinction of $A_{V} \leq 0.20$; requiring either a detection or
non-detection in $NUV$;  requiring a detection in $v_{SM}$; requiring
an inferred distance of less than 500 parsecs; and finally, requiring
an age precision of better than 7\%.

We also comment on the ends of our parameter space. First, the
fact that we barely detect any stars with $\tau/\rm{Gyr} \leq
2$ is due to our color-magnitude selection function (see Figure
\ref{fig:selection}). Second, the lack of stars with [Fe/H] $> 0.50$
is due to that being the end of the MIST isochrone grid, and likely
as well the fact that those stars are intrinsically rare. Finally, we
have a reassuringly small excess of stars with $\tau/\rm{Gyr} \approx
13.7$. That is a cosmologically-motivated end to our allowed parameter
space, nevertheless, we would expect to see a large pileup of stars
at old ages if either of our measurement or methodological errors were
larger, for example if we frequently and significantly underestimated
reddening or metallicity. Though there is a pileup that can be seen at
the right end of the left panel of Figure \ref{fig:AgeMetDistributions},
it is reassuringly small.

\begin{figure}
\centering
\includegraphics[width=0.47\textwidth]{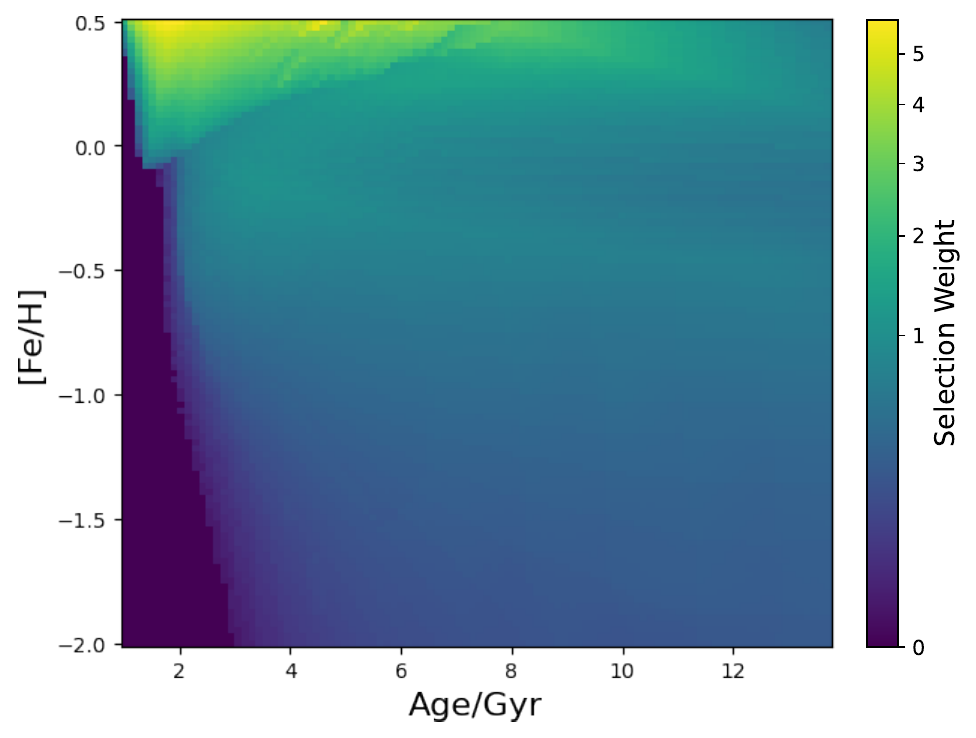}
\caption{An approximation of our selection function as a function of age
and metallicity, where the selection weight is normalized to 1 for a star
of solar age and metallicity. The selection function is close to uniform
in the age-metallicity regime where stars are found in numerous numbers. }
\label{fig:Selection}
\end{figure}

We estimate our selection bias as follows. First, we require the stars to
have predicted photometric parameters satisfying those in the top panel
of Figure \ref{fig:selection}. We then require that the star either have
a predicted $14 \leq NUV \leq 22$,  $12 \leq u_{\rm{SDSS}} \leq 17$,
or   $11 \leq u_{\rm{SM}} \leq 17$, as is the case for the stars in our
sample, given a volumetrically uniform distribution satisfying $250 \leq
\rm{d/pc} \leq 1000$. This approximation of the selection probability,
shown in Figure \ref{fig:Selection}, varies little for stars with   $2
\leq \tau/\rm{Gyr} \leq 12$ and $-1.0 \leq \rm{[Fe/H]} \leq +0.50$. As
these stars comprise the vast majority of our sample,the selection
function has little effect on our final sample.

There are certainly factors not included in the estimate of our selection
function that could significantly adjust the selection probability. These
include a possible age and metallicity dependence to the detectable binary
fraction, the correlations between the spatial selection functions of
GALEX, SDSS, and Skymapper and the metallicity distribution functions
of field stars, and the correlation between high-extinction sightlines
(which tend to be along the Galactic plane) and the age-metallicity
distributions of those sightlines. The computation of these adjustment
factors would constitute a large study of their own, and are thus beyond
the scope of this investigation.

\begin{table*}
\caption{Summary input data and derived stellar parameters for 401,819 stars. }
\begin{tabular}{|l l l|}
\hline
\hline
Num & Label & Explanations \\
\hline
1 & dr3\_source\_id &  Unique source identifier in Gaia DR3 \citep{2023AnA...674A...1G}, for all sample stars  \\
2 & objid\_GALEX &  Unique source identifier in GALEX \citep{2017ApJS..230...24B} if in catalog, otherwise null. \\
3 & ra &  Right ascension from Gaia DR3, Epoch J2016 \\
4 & dec &  Declination from Gaia DR3, Epoch J2016 \\
5 & ra2000 &  Right ascension from Gaia DR3, Epoch J2000 \\
6 & dec2000 &  Declination from Gaia DR3, Epoch J2000 \\
7 & l2000 &   Galactic longitude from Gaia DR3, Epoch J2000 \\
8 & b2000 &  Galactic latitude from Gaia DR3, Epoch J2000 \\
9 & phot\_g\_mean\_mag &  G-band mean magnitude
 from Gaia DR3 \\
10 & mass\_16 & 16th percentile of the posterior distribution of the initial stellar mass \\
11 & mass & 50th percentile of the posterior distribution of the initial stellar mass  \\
12 & mass\_84 & 84th percentile of the posterior distribution of the initial stellar mass  \\
13 & age\_16 & 16th percentile of the posterior distribution of the stellar age \\
14 & age & 50th percentile of the posterior distribution of the stellar age   \\
15 & age\_84 & 84th percentile of the posterior distribution of the stellar age \\
16 & feh\_16 & 16th percentile of the posterior distribution of the stellar metallicity \\
17 & feh & 50th percentile of the posterior distribution of the stellar metallicity  \\
18 & feh\_84 & 84th percentile of the posterior distribution of the stellar metallicity  \\
19 & AV\_model\_16 & 16th percentile of the posterior distribution of the extinction to the star $A_{V}$ \\
20 & AV\_model & 50th percentile of the posterior distribution of the extinction to the star $A_{V}$ \\
21 & AV\_model\_84 & 84th percentile of the posterior distribution of the extinction to the star $A_{V}$ \\
22 & Teff\_16 & 16th percentile of the posterior distribution of the stellar surface effective temperature $T_{\rm{eff}}$ \\
23 & Teff & 50th percentile of the posterior distribution of the stellar surface effective temperature $T_{\rm{eff}}$ \\
24 & Teff\_84 & 84th percentile of the posterior distribution of the stellar surface effective temperature $T_{\rm{eff}}$ \\
25 & logg\_16 & 16th percentile of the posterior distribution of the stellar surface gravity $\log{g}$ \\
26 & logg & 50th percentile of the posterior distribution of the stellar surface gravity $\log{g}$  \\
27 & logg\_84 & 84th percentile of the posterior distribution of the stellar surface gravity $\log{g}$  \\
28 & GP\_Jp &  Azimuthal action, all dynamical variables derived by GalPy \citep{2015ApJS..216...29B}. \\
29 & GP\_Jr &  Radial action. \\
30 & GP\_Jz & Vertical action.  \\
31 & GP\_Lz & Orbital angular momentum about the Galactic major axis. \\
32 & GP\_E & Energy of orbit, negative for bound orbits. \\
33 & GP\_Rperi & Smallest distance to galactic center of orbit \\
34 & GP\_Rap & Largest distance to galactic center of orbit \\
35 & GP\_zmax & Largest separation from the plane of orbit. \\
36 & GP\_ecc & Eccentricity of orbit. \\
37 & Blend &  ``1" for stars for which photometry appears blended as per the criteria of Section \ref{subsec:MorePhot}, ``0" otherwise. \\
38 & Chance\_pvalue & $p$-value from \citet{2022arXiv220611275C} if measured, otherwise null.  \\
39 & VarExcess & Percentile of photometric variability in Gaia G-band with respect to the rest of the sample. \\
40 & Sample & ``PS" for primary sample, ``MPannex" for metal-poor annex \\
\hline
\end{tabular}
\label{Table:FullDataPrime}
\tablecomments{Summary input data and derived parameter for
401,819 subgiants.  This table is published in its entirety in the
machine-readable format in the online edition.}
\end{table*}

\subsection{The Associations Between Derived Ages and Dynamics}

We estimate the dynamical properties of our sample stars using
\texttt{galpy\footnote{\url{http://github.com/jobovy/galpy}}}
\citep{2015ApJS..216...29B}, and compute the orbits using the module
developed by \citet{2018PASP..130k4501M}. Some of our results may
be inverted relative to other studies as galpy uses a left-handed
Galactocentric coordinate frame.

We adopted the \texttt{MWPotential2014} described by
\citet{2015ApJS..216...29B}. In that model, the bulge is parameterized
as a power-law density profile that is exponentially cut-off at 1.9
kpc with a powerlaw exponent of $-$1.8. The disk is represented by
a Miyamoto-Nagai potential with a radial scale length of 3kpc and a
vertical scale height of 280 pc \citep{1975PASJ...27..533M}. The halo
is modeled as a Navarro–Frenk–White halo with a scale length of
16kpc \citep{1996ApJ...462..563N}. We set the solar distance to the
Galactic center to $R_{0}=8.122$ kpc \citep{2018A&A...615L..15G},
the circular velocity at the Sun to $V_{0}=238$km/s
\citep{2012MNRAS.427..274S,2016ARA&A..54..529B}, the height of the Sun
above the plane to $z_{0}= 25$ pc, and the solar motion with the respect
to the local standard of rest to $(U_{\odot},V_{\odot},W_{\odot})=(10.0,
11.0,7.0)$ km/s \citep{2008ApJ...673..864J}, where the latter is
consistent with the values of $(U_{\odot},V_{\odot},W_{\odot})=(11, 12,
7)$ km/s derived by \citet{2010MNRAS.403.1829S}.

\begin{figure*}
\centering
\includegraphics[width=0.80\textwidth]{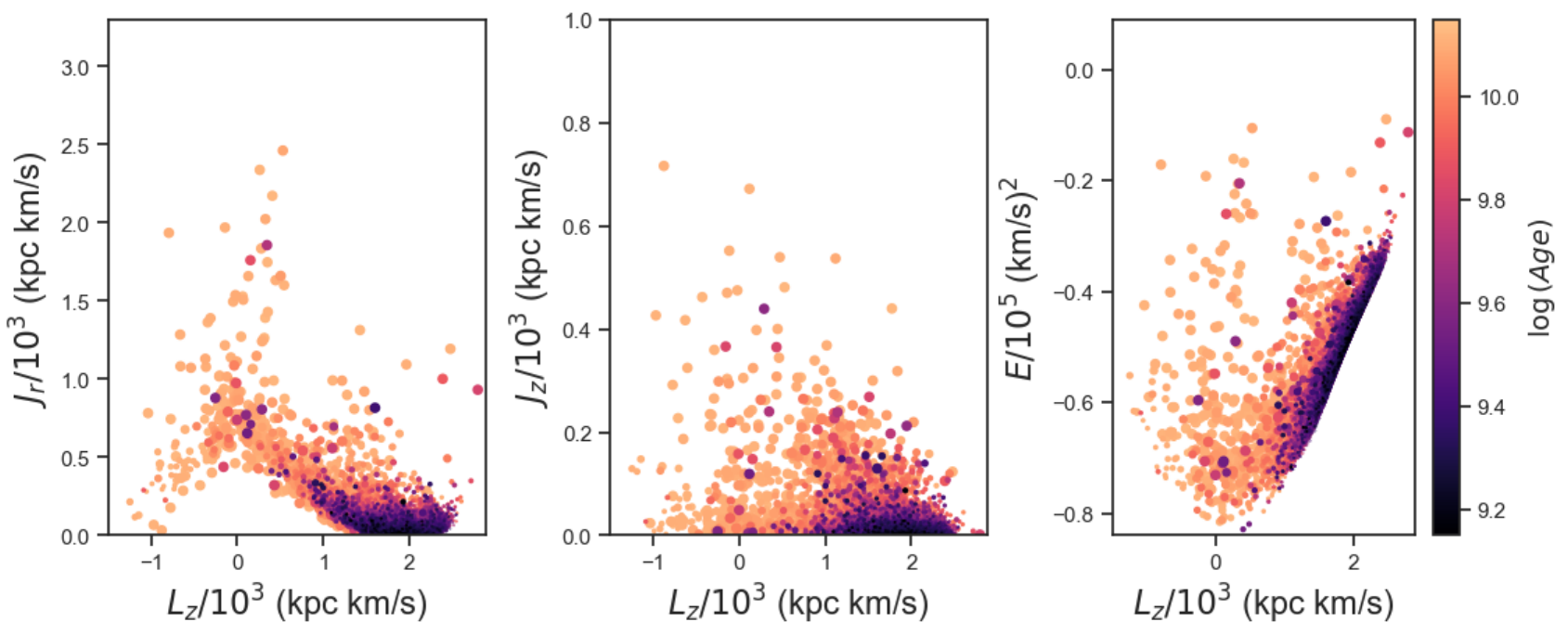}
\caption{Older stars can be found in any populated region of dynamic
phase space, but younger stars are virtually purely associated with
cold disk kinematics.  The dynamical distribution of our sample stars
in action space (Left two panels) and in terms of integrals of motion
(right panel). The points are color-coded by derived age, and the size
of the points is proportional to their isolation in $J_{r}-J_{z}$ space.}
\label{fig:MainSampleDynamics}
\end{figure*}

\begin{figure*}
\centering
\includegraphics[width=0.80\textwidth]{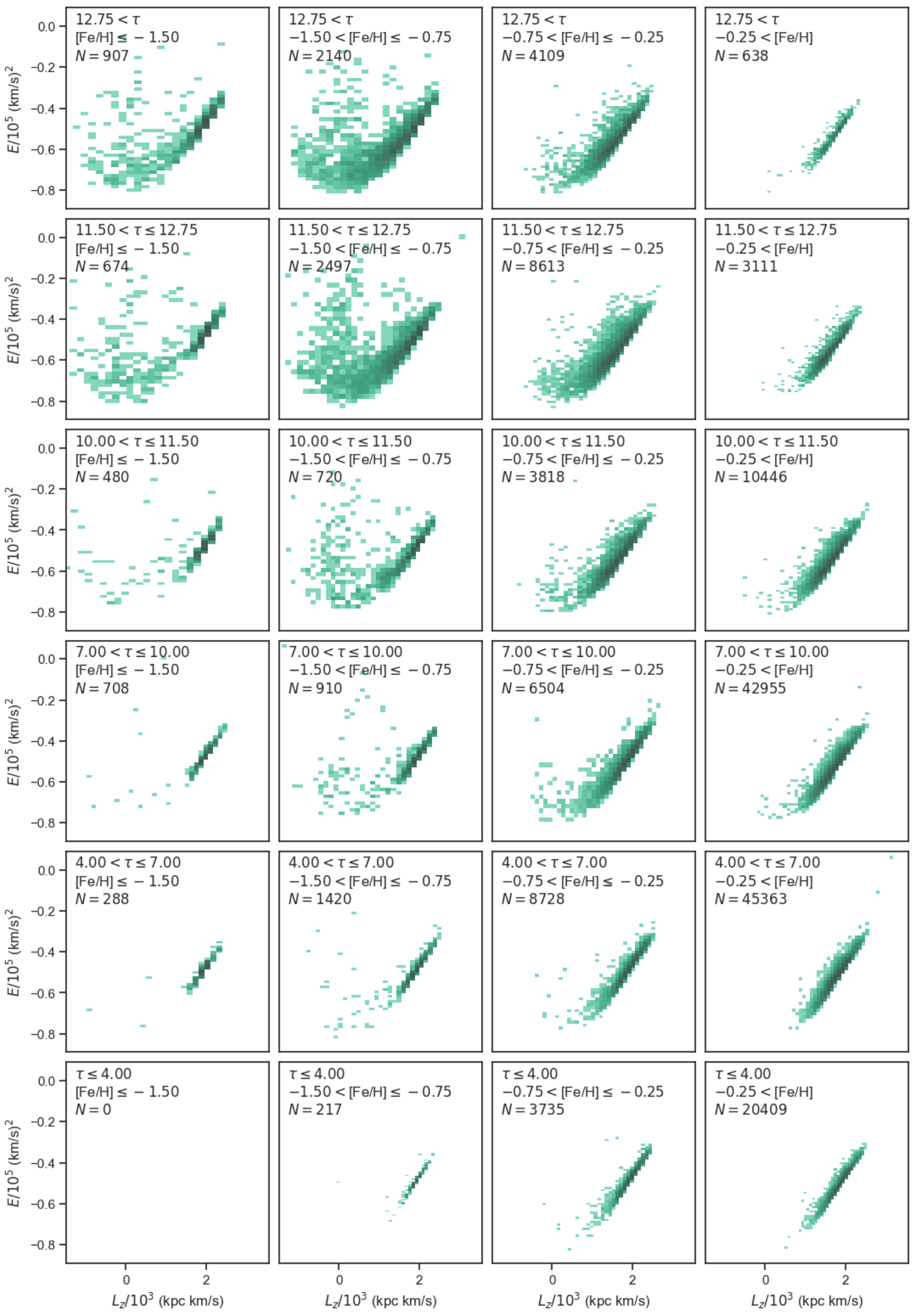}
\caption{The derived $E$ vs $L_{Z}$ distribution functions for the
orbits of the stars in our combined primary + metal-poor annex sample,
in six different bins of age (increasing upwards) and four different bins
of metallicity (increasing to the right) for 24 total bins. We include
169,411 stars that satisfy our variability and blending criteria, for
which the 1-$\sigma$ precision in the derived ages is better than 10\%,
and for which the  1-$\sigma$ precision in derived [Fe/H] is better
than 0.20 dex. Younger stars have nearly entirely disk-like kinematics,
whereas older stars can be found in any region of kinematic phase space
where there are stars to be found, including the Disk.}
\label{fig:MainSampleDynamics2}
\end{figure*}

In Figure \ref{fig:MainSampleDynamics}, we show the distribution of
estimated orbital parameters for 105,445 sample stars with derived ages
that are estimated to be precise to 7\% or better. The distribution in
action space is shown in the two left columns, and in the right column
we show the distribution as a function of two integrals of motion
(energy and angular momentum around the Galaxy's rotation axis). As
our assumed gravitational potential for the Milky Way is axisymmetric,
the action $J_{\phi}$ is exactly equal to the angular momentum $L_{z}$.
Then, in Figure \ref{fig:MainSampleDynamics2}, we show the distribution
of stars of our combined sample (including the metal-poor annex) as a
function of $E$ vs $L_{z}$, binned by age and metallicity.

What we find is that the older stars are broadly distributed
in dynamical space, that older stars can in fact be found in any
region of dynamical phase space where we find any stars at all. The
trend is for the dynamical distribution of stars to become more
and more localized as the derived ages of stars become younger and
younger. This result is qualitatively consistent with the Milky Way
Disk formation model of \citet{2009MNRAS.399.1145S}, which postulated
that the thick and thin disks were born at the same time. That is in
contrast to various other observational arguments that the oldest
Thick Disk are several Gyr older than the oldest Thin Disk stars
(e.g. \citealt{2008MNRAS.388.1175H,2017ApJ...837..162K}). There
is evidence previously discussed in the literature of ancient
stars on Thin Disk orbits, but these have thus far been few and
far between \citep{2011ApJ...737....9R,2018ApJ...867...98S},
though this subject appears to be on the ascendant
\citep{2023arXiv231102297H,2023arXiv231109294Z,2023arXiv231202356B}. To
the best of our knowledge, ours is the first analysis of a sample of
these stars for which precise ages are available.

\subsection{The Metal-Poor Annex} \label{subsec:MPannex}

We constructed the metal-poor annex, using the criteria delineated in
Section \ref{sec:SampleSelection}, so as to extract a sample of metal-poor
subgiant stars. These are indistinguishable from the vastly more numerous
metal-rich turnoff stars on the optical color-magnitude diagram, but can
be selected statistically using a combination of ultraviolet and optical
colors, as can be better understood by inspecting Figure \ref{fig:GALAH}.

The metal-poor annex totals 112,062 stars, of which 96,247 stars meet
the photometric and blending criteria used for the main sample  --
that is the distribution that we plot in the right panel of Figure
\ref{fig:AgeMetDistributions}. Of those, 14,670 stars have derived age
precision better than 7\%, and 6,768 have derived age precisions better
than 5\%.

\begin{figure}
\centering
\includegraphics[width=0.45\textwidth]{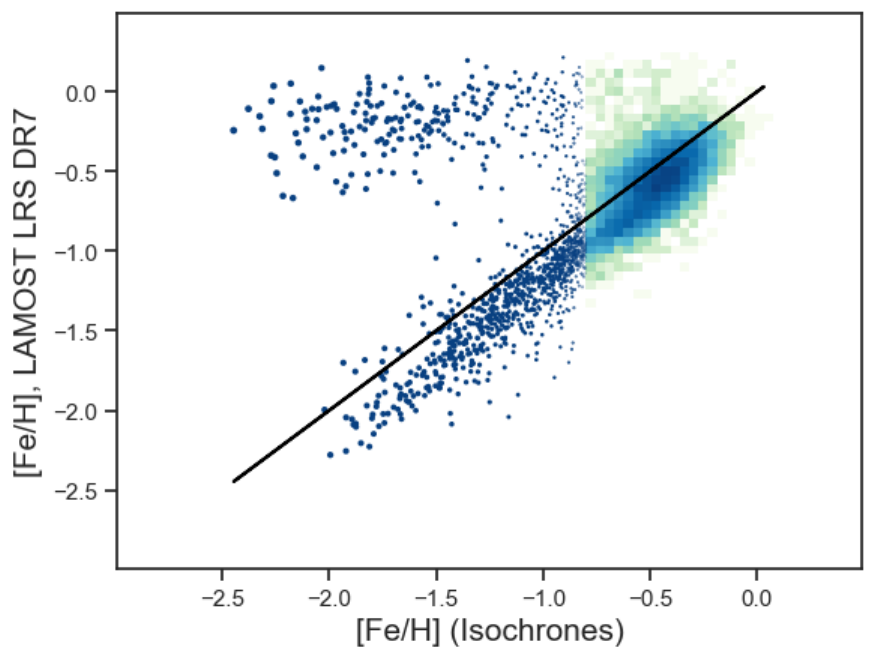}
\caption{For stars in our metal-poor annex that have been observed
by LAMOST. The cloud of stars for which the photometric metallicities
are under-estimated are largely a product of less reliable ultraviolet
photometry and less certain reddening measurements. For the vast majority
of stars, the astrophotometric metallicities are generally consistent
with the spectroscopically-derived metallicities }
\label{MP_LAMOST_Comparison}
\end{figure}

Of the 96,247 metal-poor annex stars that meet our photometric and
variability criteria, 10,015 have matches in LAMOST DR7, and we show
their distribution in Figure \ref{MP_LAMOST_Comparison}. The two samples
are consistent, with a median offset of 0.11 dex and a median absolute
deviation of 0.10 dex. As with Figure \ref{fig:MetComparisons}, we do
see a cloud of points for which the spectroscopic metallicities exceed
the photometric metallicities. Here, we find that the error is due to
specific failure modes of the photometric analysis. Some 30\% of the
stars in Figure \ref{MP_LAMOST_Comparison} with derived metallicities
satisfying [Fe/H] $\leq -1.0$ have a difference in their photometric
and spectroscopic metallicities exceeding 0.50 dex. For those, the
mean derived ages are 24\% lower, the mean derived extinctions $\gtrsim
2\times$ greater, and the mean derived errors on the metallicities are
$\gtrsim 5\times$ greater.

The most significant offset, however, is that due to whether or
not there is an $NUV$ measurement. For the 209 stars in Figure
\ref{MP_LAMOST_Comparison} with derived metallicities satisfying [Fe/H]
$\leq -1.0$ that have an offset between the photometric and spectroscopic
metallicity determinations exceeding $\Delta$[Fe/H]$=$0.50 dex, only
1 has an $NUV$ measurement. In contrast, for the 486 other stars
with derived metallicities satisfying [Fe/H] $\leq -1.0$, 399 have
$NUV$ measurements.  We conclude that $NUV$ is simply more effective
than $u$-band at distinguishing metal-poor subgiant stars from young
metal-rich turnoff stars with similar optical colors -- particularly
when the reddening is high.

The metal-poor annex yields significant potential for follow-up. Among
the stars that satisfy our blending and variability criteria and have
7\% precision in their age determinations, and that are brighter than
$G=14$, some 4,649 have [Fe/H] $\leq-1.00$, and some 3,123 have [Fe/H]
$\leq -1.5$. Of these, roughly half are brighter than $G=13$, further
enabling follow-up study.

\subsubsection{Metal-Poor Annex Systematics and Uncertainties}
\label{syst_unc_mp}

We investigated the fidelity of our photospheric stellar parameters
for the metal-poor annex by performing spectroscopic observations of
apparently metal-poor subgiants (i.e., $[\text{Fe/H}] \lesssim -1.0$),
one young and one old.  To remain as unbiased as possible, we selected
stars observable at the time and from the place of observation based
on their inferred metallicities and ages alone.  We selected Gaia DR3
6032690864784481664 ($T_{\text{eff}} = 6430 \pm 50$ K, $\log{g} = 3.90
\pm 0.02$, $[\text{Fe/H}] = -1.13 \pm 0.13$, and $\tau = 6.3 \pm 0.4$
Gyr) and Gaia DR3 6310653893928413696 ($T_{\text{eff}} = 6010 \pm 60$ K,
$\log{g} = 4.02 \pm 0.02$, $[\text{Fe/H}] = -0.82 \pm 0.08$, and $\tau =
11.2 \pm 0.8$ Gyr).

We collected their spectra with the Magellan Inamori Kyocera
Echelle (MIKE) spectrograph on the Magellan Clay Telescope at Las
Campanas Observatory \citep{bernstein2003,shectman2003}.  We used the
0\farcs7~slit with standard blue and red grating azimuths, yielding
spectra between 335 nm and 950 nm with resolution $R \approx40,\!000$
in the blue and $R \approx 31,\!000$ in the red arms.  We collected
all calibration data (e.g., bias, quartz \& ``milky" flat field, and
ThAr lamp frames) in the afternoon before each night of observations.
We reduced the raw spectra and calibration frames using the
\texttt{CarPy}\footnote{\url{http://code.obs.carnegiescience.edu/mike}}
software package \citep{kelson2000,kelson2003,kelson2014}.
We placed the spectra in the rest frame and
continuum-normalized them using Spectroscopy Made Harder
\citep[\texttt{smhr};][]{casey2014}.\footnote{\url{https://github.com/andycasey/smhr/tree/py38-mpl313}}

We analyzed Gaia DR3 6032690864784481664 and Gaia
DR3 36310653893928413696 using the methodology described in
\citet{reggiani2021,reggiani2022b,reggiani2022a,reggiani2023,2024AJ....167...45R}.
That methodology initially uses the classical
excitation/ionization/reduced equivalent width balance approach to infer
an initial set of photospheric stellar parameters.  Those parameters are
then included in the likelihood of an \texttt{isochrones} analysis that
in all other details corresponds to the analyses described in Section
\ref{sec:Clusters}.  We next use ($T_{\text{eff}}$, $\log{g}$) samples
from the resulting \texttt{isochrones} posteriors as constraints on the
spectroscopic analysis and recalculate [Fe/H] using reduced equivalent
width balance.  The photospheric stellar parameters are again included in
an \texttt{isochrones} analysis.  This procedure is iterated a few times
until the photospheric stellar parameters have converged.  Using this
approach, for Gaia DR3 6032690864784481664 we find $T_{\text{eff}} =
6310 \pm 50$ K, $\log{g} = 3.99 \pm 0.02$, $[\text{Fe/H}] = -0.44 \pm
0.10$, and $\tau = 4.8 \pm 0.3$ Gyr; for Gaia DR3 6310653893928413696
we find $T_{\text{eff}} = 6030 \pm 10$ K, $\log{g} = 3.96 \pm 0.01$,
$[\text{Fe/H}] = -1.21 \pm 0.02$, and $\tau = 13.4 \pm 0.1$ Gyr.

While our spectroscopy-based metallicity inference for Gaia DR3
6032690864784481664 is much higher than the metallicity returned by
our default analysis, the age inferences from both approaches agree
that the star is mature.  Likewise, the $T_{\text{eff}}$ and $\log{g}$
inferences from both approaches agree to within 120 K and 0.1 dex.
While our spectroscopy-based metallicity inference for Gaia DR3
6310653893928413696 is marginally lower than the metallicity returned
by our default analysis, the $T_{\text{eff}}$ and $\log{g}$ values are
in excellent agreement.  Both our default and spectroscopic analyses
agree that Gaia DR3 6310653893928413696 is ancient.

As we argued above, GALEX data is a necessary ingredient for our
highest-quality stellar parameter inferences.  GALEX data is unavailable
for both Gaia DR3 6032690864784481664 and Gaia DR3 6310653893928413696
though, so the spectroscopic results presented here represent a worst-case
scenario for the accuracy of our metal-poor stellar parameter inferences.
Gaia DR3 6032690864784481664 has $T_{\text{eff}} \approx 6400$ K placing
it at the warm end of our subgiant $T_{\text{eff}}$ distribution, so
it seems that GALEX photometry is especially important for accurate
metallicity inferences for relatively warm subgiants.

\subsection{On the Uncertainties in the Derived Ages}

\begin{figure}
\centering
\includegraphics[width=0.45\textwidth]{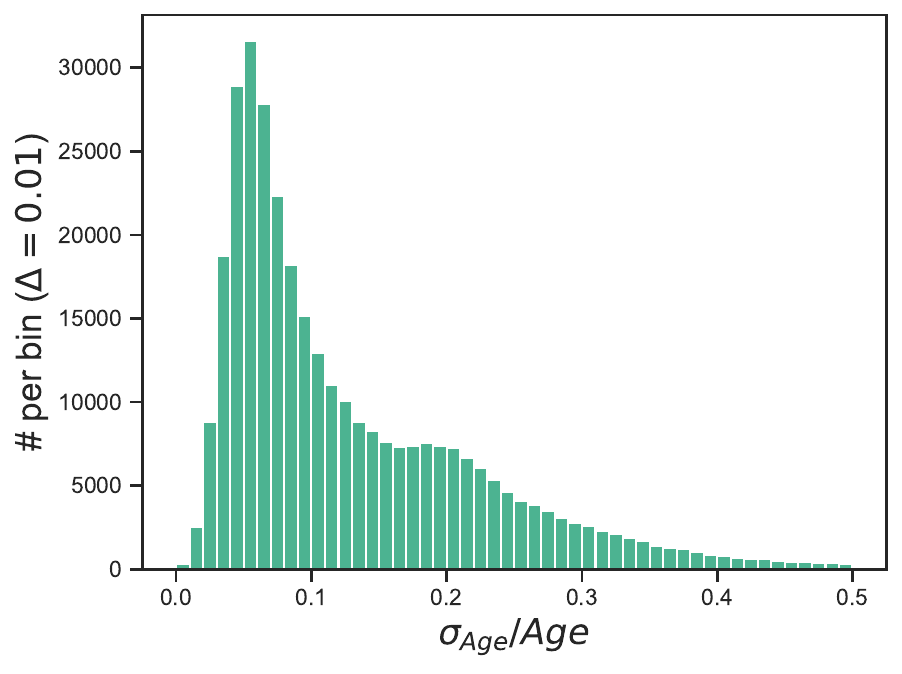}
\caption{Distribution of derived age precision for 335,778 stars in our combined primary sample and metal-poor annex that meet our quality inclusion criteria for variability and cross-matching between surveys. The median age precision is 9.5\% and the mean age precision of 13.4\%, with a long tail to higher errors. }
\label{Fig:AgeErrorHistogram}.
\end{figure}

As delineated in Equation \ref{EQ:precision} and can be inferred from
Figure \ref{fig:selection}, a 1\% precision in the parallaxes should
correspond to a 2\% precision in the derived ages in the best-case scenario where stellar temperature and metallicity are precisely determined. It thus needs to
be better understood why, in spite of a median parallax precision
of $\pi/\sigma_{\pi}=80$. It thus needs to be better understood why, in spite of a median parallax precision of $\pi/\sigma_{\pi}=80$, we achieve, for the 335,778 stars in our combined primary sample and metal-poor annex that meet our quality inclusion criteria, a median age precision of 9.5\% and a mean age precision of 13.4\% (distribution shown in Figure \ref{Fig:AgeErrorHistogram}). We note that these values drop to 8.4\% and 12.7\% respectively for stars with GALEX photometry, indicative of the greater diagnostic power of space-based, ultraviolet photometry.

Here we discuss two factors that are exacerbating the uncertainties in
derived ages, and their prospects for improvement.

The first factor is the degeneracy between the uncertainties in metallicity and age. The average correlation coefficient between the uncertainties in $\log{\rm{Age}}$ and [Fe/H] is $-0.72$. Thus, the error in metallicity, which can be largely attributed to uncertainties in the ultraviolet photometry and in the 3D extinction maps, is responsible for most of the error in the derived ages. This problem will be mitigated in the near future as the availability of robust ultraviolet photometry is expected to both increase and improve.

The first factor is the degeneracy between the uncertainties in
metallicity and age. The average correlation coefficient between the
uncertainties in $\log{\rm{Age}}$ and [Fe/H] is $-0.72$. Thus, the error
in metallicity, which can be largely attributed to uncertainties in the
ultraviolet photometry and in the 3D extinction maps, is responsible for
most of the error in the derived ages. This problem will be mitigated
in the near future as the availability of robust ultraviolet photometry
is expected to both increase and improve.

\begin{figure*}
\centering
\includegraphics[width=0.90\textwidth]{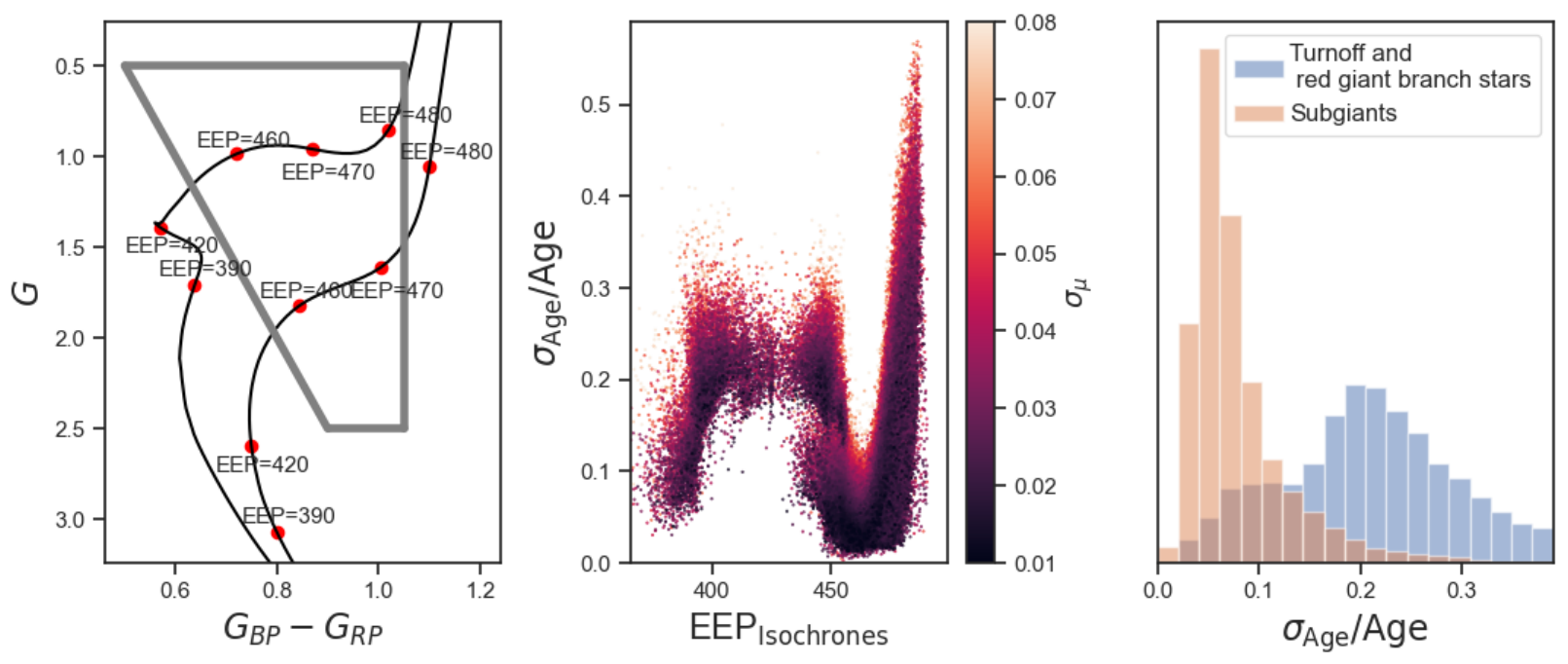}
\caption{LEFT: MIST isochrones of solar metallicity and of ages $\tau=3,8$
Gyr, with indicative EEPs shown as red points. MIDDLE: The uncertainty
in derived ages as a function of EEP, color-coded by the uncertainty
in apparent distance modulus. RIGHT: Normalized distribution of age
precisions for stars with $450 \leq \rm{EEP} \leq 475$ (subgiants, orange)
and other stars (blue). We see that derived ages are more precise at all
phases of stellar evolution when the assumed apparent distance moduli are
more precise, and the best precision is achieved on the subgiant branch.}
\label{Fig:EEP}.
\end{figure*}

To further explore the origin of the uncertainties in derived ages,
we define a coarse estimate of the uncertainty in apparent distance
modulus $\sigma_{\mu}$, as follows:
\begin{equation}
\sigma_{\mu}^2 = \biggl( \frac{5}{\ln{10}}\frac{\sigma_{\pi}}{\pi} \biggl)^2 + \sigma_{AV}^2.
\end{equation}

In the middle panel of Figure \ref{Fig:EEP}, we show that derived ages are
indeed more precise for stars with smaller values of $\sigma_{\mu}$, but
that another significant contributor to the uncertainty in derived ages is
that of the derived EEP (equivalent evolutionary point) of the star. That
is because our selection function inevitably includes not only subgiant
branch stars, but also main-sequence turnoff and first-ascent red giant
branch stars, as shown in the left panel. Returning to the middle panel,
we see that we do achieve excellent precision in derived ages for stars
with EEP values of around 460, corresponding to subgiant branch stars. The
errors are much higher for those stars which are either turnoff stars (EEP
$\approx$ 420) or first-ascent red giant branch stars (EEP $\gtrapprox$
475). In our case, some 75\% of the stars in our sample have derived EEPs
in the range ($450 \leq \rm{EEP} \leq 475$) approximately corresponding
to the subgiant branch. On the left panel, we show that the normalized
distribution in derived ages for subgiant stars, as approximated by
those stars with ($450 \leq \rm{EEP} \leq 475$), has a mode at around
5\%. A selection function that more tightly focused on these stars could
likely be constructed by making a finer use of ultraviolet photometry,
and to make the color-magnitude selection box more metallicity dependent.

\section{Summary, Discussion, and Conclusion} \label{sec:Conclusion}

In this investigation, we have demonstrated that the combination of
available data for parallaxes, ultraviolet-through-infrared photometry,
variability, and three-dimensional extinction maps are adequate to
obtain precise ages and metallicities for nearly half-a-million subgiant
stars. This is, among other successes, an indicator of the spectacular
triumph within astronomy of each of improving observations and related
technology, increasingly sophisticated computational and data analysis
methods, and improving stellar evolution models. Our metal-poor annex
is also, arguably, among the largest samples available for detailed
investigative and spectroscopic follow-up of the earliest phases of
Milky Way formation and assembly. Below, we discuss some of the larger
sources of uncertainty in our results, prospects for improved analysis
in the future,

\subsection{Major Systematic Uncertainties: Variability, Elemental
Abundance Variations, and Bolometric Corrections}

Undiagnosed binaries, products of binary evolution, and other variables,
are plausibly the most significant source of systematic error in our
sample. As can be seen in Figure \ref{fig:Clusters}, our methods identify
some, but not all such stars on the subgiant branches of the open clusters
M67 and NGC 188. This may be an over-representation of this error source,
as these clusters are both more distant than the majority of our sample,
and so the binary diagnostic criteria may be less reliable. The issue may
also mitigate over time -- we are eager to find out how many additional
variables will be identified once Gaia DR4 is released, as the astrometric
and photometric time-series will cover 66 rather than 34 months of data,
and the individual measurements in the radial velocity measurements will
become available.

A second systematic uncertainty is that our analysis has assumed
the scaled-solar composition isochrones available from the MIST
database. We have done so as these are the stellar models used by
the \texttt{isochrones} package. The most obvious uncertainty
here is that the assumption of scaled-solar composition is
not valid for stars of the thick disk, halo, and many accreted
streams (and thus most metal-poor stars), as these stars tend to be
enhanced in the $\alpha$-elements (O, Ne, Mg, Si, S, Ar, Ca, Ti, see e.g.
\citealt{1991AJ....102.2001S,1994ApJS...91..749M,2010A&A...513A..35A,2014A&A...562A..71B,2015ApJ...808..132H}),
and often also have variations in carbon and nitrogen abundances
\citep{2015MNRAS.453.1855M,2021MNRAS.500.5462H}, and possibly helium
\citep{2007MNRAS.382.1516C,2010A&A...518A..13G}. These abundance changes
will effect the evolutionary state, temperature, luminosity, intrinsic
colors of stars of otherwise identical mass and age. This is an issue
that should resolve itself in the future, as the next generation of MIST
isochrones will include stellar models of variable $\alpha$-abundance
(Aaron Dotter, private communication).

\subsection{Major Quantitative Limitation: The Availability of
Ultraviolet-Through-Infrared Photometry for Stars With Precise
Parallaxes.}

Our optical, Gaia-based selection query for candidate subgiant
stars in our primary sample, which is shown graphically in Figure
\ref{fig:selection} and for which we provide the ADQL query in the
appendix, yields 1,998,909 sources. Once we require unambiguous matches
between Gaia DR2 and Gaia DR3 and full photometry from 2MASS and WISE,
we are left with  1,661,347 matches. Once we require reddening data, and
correct the color-magnitude selection box in Figure \ref{fig:selection}
to be based on estimated values of $(G_{\rm{BP}} - G_{\rm{BP}})_{0}$,
we are left with a sample of 618,598 stars.  Once we then require $A_{V}
\leq 0.50$, the sample drops to 462,045 stars. Once we also require an
ultraviolet photometric measurement, the sample drops to 289,718 stars.

These numbers will be shifted upwards by a significant amount with
future data. For example, Gaia DR4 and Gaia DR5 will respectively cover
an expected 66 and 120 months of data, resulting in parallaxes that are
respectively two and four times more precise than those of Gaia DR3. We
should eventually be able to sample stars at distances that are up to
four times further away, enabling the initial sample to exceed 10$^7$
subgiant stars. At the ultraviolet end, vastly more data will be available
in the future by missions such as ULTRASAT \citep{2023arXiv230414482S},
UVEX \citep{2021arXiv211115608K} and the Vera Rubin Observatory
\citep{2019ApJ...873..111I}. The first two will provide in measurements
in two bands similar to those of GALEX but down to 23rd and 25th
magnitudes respectively, and the third will provide $u$-band photometry
for three quarters of the sky down to 26th magnitude. For each of these,
it is expected that the PSF will be sharper and that the absolute flux
calibration will be more accurate than currently available ultraviolet
data. At the near-infrared end, large parts of the sky will be observed
to 10 or more magnitudes deeper than 2MASS, and with a 5-10 $\times$
smaller PSF, by Euclid \citep{2011arXiv1110.3193L} and the Nancy Grace
Roman Space Telescope \citep{2015arXiv150303757S}.

\subsection{Major Improvements Feasible With Currently Available Data}

Our aim with this project was both to develop a catalog and to develop
a methodology. Thus, our use of publicly-available data was extensive,
but it was not exhaustive.

Four large photometric surveys that we did not make use are the Southern
Photometric Local Universe Survey (S-PLUS, \citealt{2019MNRAS.489..241M}),
the VST Photometric H$\alpha$ Survey of the Southern Galactic Plane
and Bulge (VPHAS$+$, \citealt{2014MNRAS.440.2036D}), The Blanco DECam
Bulge Survey (BDBS, \citealt{2020MNRAS.499.2357J}), and the DeCam Plane
Survey (DeCaPs, \citealt{2018ApJS..234...39S}). The first three have
ultraviolet and optical photometry for large parts of the sky down to
approximately 20th magnitude, and the fourth has optical grizY photometry
for approximately 2 billion point sources across the Galactic plane. These
datasets would undoubtedly significantly increase our sample size,
and are obvious candidates for inclusion in any follow-up study.

Another option with the possibility of qualitatively enhancing
the analysis in this work is the inclusion of the spectra from Gaia,
specifically the $BP/RP$ spectra \citep{2021A&A...652A..86C} and that from
the radial velocity spectrometer (RVS, \citealt{2019A&A...622A.205K}). The
former covers the wavelength range 300–10,500 \AA\, with a resolution
between 13 and 85, and the latter covers the wavelength range 8450-8720
\AA\, with a resolution of 11,500.

In our Figure \ref{fig:MetComparisons}, we show that metallicity
measurements from the RVS are reliable down to [Fe/H] $\approx -0.50$,
and this was also shown to be the case for the $BP/RP$ spectra by
\citet{2022MNRAS.516.3254W}, see their Figure 4. However, in both cases
the limitations of these spectra, likely due to a degeneracy between
metallicity and temperature in the regime of weaker absorption lines,
have been quantified when the spectra are used to determine stellar
parameters \textit{on their own}. The diagnostic power would undoubtedly
be greatly enhanced once one includes information from parallaxes,
precise photometry in the ultraviolet and infrared, and constraints from
published three-dimensional extinction maps.

\subsection{Conclusions}

This study of astrophotometric age and metallicity determinations for
solar neighborhood subgiant stars represents one of the largest catalogs
of stars with precisely measured relative ages, for which the distribution
is shown in our Figure \ref{fig:AgeMetDistributions}.

We have validated our analysis using several different independent and
complementary methods. We showed, via comparisons with open clusters,
that we can infer precise relative ages; and we showed, via our comparison
to large spectroscopic surveys (APOGEE, GALAH, and LAMOST) shows that we
can infer consistently accurate metallicities across the interval $-2.0
\leq \rm{[Fe/H]} \leq +0.50$ to a precision no worse than $\sim$0.10
dex, and that includes the effect of systematics such as undiagnosed
binaries. We have also shown that the median systematic error in the
bolometric zero points is approximately 0.01 mag, and that our analysis
is insensitive to small changes in the input data and priors.

Our dynamical analysis, shown in Figures \ref{fig:MainSampleDynamics}
and \ref{fig:MainSampleDynamics2}, shows that older stars are to be found
in any region of dynamic phase space where there are stars of any age to
be found, but that younger stars are progressively more constrained to
colder regions of dynamic phase space. The range of values of $J_{r}$
and $J_{z}$ drops to being closer and closer to zero as ages drop,
and the orbits become more tightly constrained to circular orbits as
per their distributions in the integrals of motion $E$ and $L_{z}$.

Our metal-poor annex is one of the largest catalogs of metal-poor stars
with precise metallicities and ages -- it includes over 4,500 stars with
both [Fe/H] $\leq -1.00$, age precisions better than 7\%, and that are
brighter than $G=14$. Of those,  3,123 stars have best-fit metallicities
satisfying [Fe/H] $\leq -1.50$.

We look forward to further vetting of this method by subsequent studies,
as well as the promising prospects to see the sample sizes increase as
future data become available, and for the precision to improve as more
data are included.

\section*{Acknowledgments}

We thank Tim Morton, Rosemary Wyse, Luca Casagrande, Sven Buder, and
Aaron Dotter for helpful discussions.  David M. Nataf acknowledges
support from NASA under award Number 80NSSC21K1570 and award
Number 80NSSC19K058. Henrique Reggiani acknowledges support
from a Carnegie Fellowship.  This research made use of Astropy
\footnote{http://www.astropy.org} a community-developed core
Python package for Astronomy, \citep{astropy:2013,astropy:2018}.
This research has made use of the WEBDA database, operated at the
Department of Theoretical Physics and Astrophysics of the Masaryk
University This work has made use of data from the European Space
Agency (ESA) mission {\it Gaia} (\url{https://www.cosmos.esa.int/gaia}),
processed by the {\it Gaia} Data Processing and Analysis Consortium (DPAC,
\url{https://www.cosmos.esa.int/web/gaia/dpac/consortium}). Funding for
the DPAC has been provided by national institutions, in particular the
institutions participating in the {\it Gaia} Multilateral Agreement.
The national facility capability for SkyMapper has been funded
through ARC LIEF grant LE130100104 from the Australian Research
Council, awarded to the University of Sydney, the Australian National
University, Swinburne University of Technology, the University of
Queensland, the University of Western Australia, the University of
Melbourne, Curtin University of Technology, Monash University and the
Australian Astronomical Observatory. SkyMapper is owned and operated
by The Australian National University's Research School of Astronomy
and Astrophysics. The survey data were processed and provided by
the SkyMapper Team at ANU. The SkyMapper node of the All-Sky Virtual
Observatory (ASVO) is hosted at the National Computational Infrastructure
(NCI). Development and support of the SkyMapper node of the ASVO has been
funded in part by Astronomy Australia Limited (AAL) and the Australian
Government through the Commonwealth's Education Investment Fund (EIF)
and National Collaborative Research Infrastructure Strategy (NCRIS),
particularly the National eResearch Collaboration Tools and Resources
(NeCTAR) and the Australian National Data Service Projects (ANDS).
This publication makes use of data products from the Two Micron All
Sky Survey, which is a joint project of the University of Massachusetts
and the Infrared Processing and Analysis Center/California Institute of
Technology, funded by the National Aeronautics and Space Administration
and the National Science Foundation.  This publication makes use of
data products from the Wide-field Infrared Survey Explorer, which is
a joint project of the University of California, Los Angeles, and the
Jet Propulsion Laboratory/California Institute of Technology, funded
by the National Aeronautics and Space Administration.  Funding for the
Sloan Digital Sky Survey IV has been provided by the Alfred P. Sloan
Foundation, the U.S. Department of Energy Office of Science, and the
Participating Institutions. SDSS-IV acknowledges support and resources
from the Center for High-Performance Computing at the University of
Utah. The SDSS web site is www.sdss.org.  SDSS-IV is managed by the
Astrophysical Research Consortium for the Participating Institutions
of the SDSS Collaboration including the Brazilian Participation Group,
the Carnegie Institution for Science, Carnegie Mellon University,
the Chilean Participation Group, the French Participation Group,
Harvard-Smithsonian Center for Astrophysics, Instituto de Astrof\'isica
de Canarias, The Johns Hopkins University, Kavli Institute for the
Physics and Mathematics of the Universe (IPMU) / University of Tokyo,
the Korean Participation Group, Lawrence Berkeley National Laboratory,
Leibniz Institut f\"ur Astrophysik Potsdam (AIP), Max-Planck-Institut
f\"ur Astronomie (MPIA Heidelberg), Max-Planck-Institut f\"ur Astrophysik
(MPA Garching), Max-Planck-Institut f\"ur Extraterrestrische Physik (MPE),
National Astronomical Observatories of China, New Mexico State University,
New York University, University of Notre Dame, Observat\'ario Nacional /
MCTI, The Ohio State University, Pennsylvania State University, Shanghai
Astronomical Observatory, United Kingdom Participation Group, Universidad
Nacional Aut\'onoma de M\'exico, University of Arizona, University
of Colorado Boulder, University of Oxford, University of Portsmouth,
University of Utah, University of Virginia, University of Washington,
University of Wisconsin, Vanderbilt University, and Yale University.

\software{Astropy \citep{astropy:2013,astropy:2018},
SciPy \citep{SciPy,scipy2020},
NumPy \citep{NumPy},
isochrones \citep{2015ascl.soft03010M},
Multinest \citep{feroz2008,feroz2009,feroz2019},
pandas \citep{scipy2010},
scipy \citep{scipy2020},
Matplotlib \citep{Hunter:2007},
TOPCAT \citep{2005ASPC..347...29T}}

\bibliography{NatafSubgiants}

\appendix
We aim to select data which are relatively robust, by making use of the
data quality flags suggested by the collaborations responsible for each
dataset, which are restated below.

\section{Photometric Data Catalog Construction and Quality Flags}
\label{subsec:PhotSelec}

\begin{enumerate}
\item
From the Gaia data archive (\url{https://gea.esac.esa.int/archive/}),
we select numerous measurements, including some that are ultimately
not used. The most significant measurements that we select are the DR3
astrometric measurements, the DR2 mean photometric measurements and
the SkyMapper DR2, SDSS DR13, 2MASS PSC, and WISE AllWISE crossmatches
identified by the Gaia collaboration. Here, we show the full query that
can be used to download a parent sample to our main sample:
\begin{verbatim}
SELECT gaiadr3.source_id AS dr3_source_id,
e.original_ext_source_id AS skymapper_id,
f.original_ext_source_id AS sdss_id,
g.original_ext_source_id AS twomass_id,
h.original_ext_source_id AS allwise_id,
i.original_ext_source_id AS panstarrs1_id,
b.dr2_source_id AS dr2_source_id,
gaiadr2.phot_g_mean_mag AS dr2_gmag,
gaiadr2.phot_bp_mean_mag AS dr2_bpmag,
gaiadr2.phot_rp_mean_mag AS dr2_rpmag,
gaiadr2.phot_bp_mean_flux_over_error as dr2_phot_bp_mean_flux_over_error,
gaiadr2.phot_g_mean_flux_over_error as dr2_phot_g_mean_flux_over_error,
gaiadr2.phot_rp_mean_flux_over_error as dr2_phot_rp_mean_flux_over_error,
gaiadr3.solution_id,
gaiadr3.random_index,
COORD1(EPOCH_PROP_POS(gaiadr3.ra, gaiadr3.dec, gaiadr3.parallax, gaiadr3.pmra,
gaiadr3.pmdec, gaiadr3.radial_velocity, gaiadr3.ref_epoch,2000)) AS ra2000,
COORD2(EPOCH_PROP_POS(gaiadr3.ra, gaiadr3.dec, gaiadr3.parallax, gaiadr3.pmra,
gaiadr3.pmdec, gaiadr3.radial_velocity, gaiadr3.ref_epoch,2000)) AS dec2000,
gaiadr3.ref_epoch,
gaiadr3.ra,
gaiadr3.dec,
gaiadr3.parallax,
gaiadr3.parallax_error,
gaiadr3.parallax_over_error,
gaiadr3.pmra,
gaiadr3.pmra_error,
gaiadr3.pmdec,
gaiadr3.pmdec_error,
gaiadr3.phot_g_n_obs,
gaiadr3.phot_g_mean_flux,
gaiadr3.phot_g_mean_flux_error,
gaiadr3.phot_g_mean_flux_over_error,
gaiadr3.phot_g_mean_mag,
gaiadr3.phot_bp_mean_mag,
gaiadr3.phot_rp_mean_mag,
gaiadr3.phot_bp_mean_flux_over_error,
gaiadr3.phot_rp_mean_flux_over_error,
gaiadr3.phot_variable_flag,
gaiadr3.l,
gaiadr3.b,
gaiadr3.astrometric_excess_noise_sig,
gaiadr3.visibility_periods_used,
gaiadr3.ruwe,
gaiadr3.astrometric_params_solved,
gaiadr3.ipd_gof_harmonic_amplitude,
gaiadr3.phot_bp_mean_flux_over_error,
gaiadr3.phot_rp_mean_flux_over_error,
gaiadr3.nu_eff_used_in_astrometry,
gaiadr3.pseudocolour,
gaiadr3.ecl_lat,
gaiadr3.radial_velocity,
gaiadr3.radial_velocity_error,
gaiadr3.grvs_mag,
gaiadr3.grvs_mag_error,
gaiadr3.rvs_spec_sig_to_noise,
d.r_lo_geo, d.r_med_geo, d.r_hi_geo,
j.teff_gspphot, j.logg_gspphot, j.mh_gspphot,
j.teff_gspspec, j.logg_gspspec, j.mh_gspspec, j.alphafe_gspspec, j.fem_gspspec,
j.mh_gspspec_lower,
j.mh_gspspec_upper,
j.flags_gspspec
FROM gaiadr3.gaia_source as gaiadr3
LEFT OUTER JOIN gaiadr3.dr2_neighbourhood b ON gaiadr3.source_id = b.dr3_source_id
LEFT OUTER JOIN gaiadr2.gaia_source gaiadr2 ON b.dr2_source_id = gaiadr2.source_id
LEFT OUTER JOIN external.gaiaedr3_distance d ON gaiadr3.source_id = d.source_id
LEFT OUTER JOIN gaiaedr3.skymapperdr2_best_neighbour e ON gaiadr3.source_id = e.source_id
LEFT OUTER JOIN gaiaedr3.sdssdr13_best_neighbour f ON gaiadr3.source_id = f.source_id
LEFT OUTER JOIN gaiaedr3.tmass_psc_xsc_best_neighbour g ON gaiadr3.source_id = g.source_id
LEFT OUTER JOIN gaiaedr3.allwise_best_neighbour h ON gaiadr3.source_id = h.source_id
LEFT OUTER JOIN gaiadr3.panstarrs1_best_neighbour i ON gaiadr3.source_id = i.source_id
LEFT OUTER JOIN gaiadr3.astrophysical_parameters j on gaiadr3.source_id = j.source_id
WHERE gaiadr3.phot_g_mean_mag - 1.90*(gaiadr3.bp_rp) - 5*LOG10(d.r_med_geo) + 5 <= 2.5
AND gaiadr3.phot_g_mean_mag - 1.90*(gaiadr3.bp_rp) - 5*LOG10(d.r_med_geo) + 5 >= 0.50
AND gaiadr3.phot_g_mean_mag - 1.90*(gaiadr3.bp_rp) - 5*LOG10(d.r_med_geo) + 5 <=
5*(gaiadr3.bp_rp-0.90)+2.5
AND gaiadr3.bp_rp <= 1.30
AND gaiadr3.duplicated_source = 'FALSE'
AND gaiadr3.phot_proc_mode = 0
AND gaiadr3.phot_variable_flag != 'VARIABLE'
AND gaiadr3.non_single_star = 0
AND gaiadr3.ruwe <= 1.4
AND gaiadr3.parallax_over_error >= 50
AND gaiadr3.ipd_gof_harmonic_amplitude <= 0.10
AND gaiadr3.astrometric_params_solved = 31
AND abs(gaiadr2.phot_g_mean_mag-gaiadr3.phot_g_mean_mag) <= 0.20
\end{verbatim}
The color-magnitude selection box for the metal-poor annex is adjusted in
the manner described in Section \ref{sec:SampleSelection}. When we lasted
the query on March 1st, 2023, it took the Gaia server 4,269 seconds to
complete, for which we got 1,998,909 matches. We then removed 902 sources
for which there were multiple Gaia DR2 sources per Gaia DR3 source.
As this query may take longer to compute in periods of heavy user use,
we recommend splitting it into sub-queries in right ascension.
\item
Corrected zeropoints to the Gaia DR3 parallaxes are applied using
the prescription of \citet{2021A&A...649A...4L}, with the source code
downloaded from \url{https://gitlab.com/icc-ub/public/gaiadr3_zeropoint}.
\item
GALEX $FUV$ and $NUV$ photometry and associated uncertainties data
are downloaded from the Revised Catalog of GALEX Ultraviolet Sources
(GUVcat\_AIS, \citealt{2017ApJS..230...24B}). We require that the centroid
of the GALEX source be within 2.5$\arcsec$ of the centroid of the Gaia
DR3 source, where we use the J2000 coordinates of the latter for the
comparison. We use the $FUV$ photometry if the parameter FUV\_artifact
(called Fafl in Vizier table) does not have bits 2 or 3, and similarly
for $NUV$ photometry and the parameter $NUV\_$artifact (called Nafl in
Vizier table). To avoid contamination from UV-bright sources that are
not the intended stars, we require that the $FUV$ and $NUV$ match be
the same, and that $(FUV - NUV) \geq 3.75$.
\item
Skymapper $uvgriz$ photometry  and associated uncertainties
\citep{2019PASA...36...33O}, for sources as delineated by the
Skymapper identifiers from the Gaia archive are downloaded from
the  dr2.master table, which is accessible from the ESO Gaia archive
(\url{https://gea.esac.esa.int/archive/}) using the following query:
\begin{spverbatim}
SELECT a.object_id, a.u_flags, a.u_nimaflags, a.u_ngood, a.u_psf, a.e_u_psf, a.v_flags, a.v_nimaflags, a.v_ngood, a.v_psf, a.e_v_psf, a.g_flags, a.g_nimaflags, a.g_ngood, a.g_psf, a.e_g_psf, a.r_flags, a.r_nimaflags, a.r_ngood, a.r_psf, a.e_r_psf, a.i_flags, a.i_nimaflags, a.i_ngood, a.i_psf, a.e_i_psf, a.z_flags, a.z_nimaflags, a.z_ngood, a.z_psf, a.e_z_psf
FROM external.skymapperdr2_master AS a
INNER JOIN user_ID.InputTable AS b ON a.object_id = b.skymapper_id
\end{spverbatim}
For each bandpass ``x", we require \code{(x\_flags=0)},
\code{(x\_nimaflags=0)} and \code{(x\_ngood>=1)}. We also
require $|G - g_{\rm{sm}} + 0.35| \leq 0.30$, to reduce the
number of spurious matches.  The Skymapper data are described at
\url{http://skymapper.anu.edu.au/data-release/}.
\item
SDSS $ugriz$ photometry and associated uncertainties
\citep{2020ApJS..249....3A}, for sources as delineated by the SDSS
identifiers from the Gaia archive, are downloaded using CASJobs
(\url{http://skyserver.sdss.org/casjobs/}) using the following query:
    \begin{spverbatim}
SELECT S.objID,
CASE WHEN ((S.flags_u & 0x10000000) != 0) AND ((S.flags_u & 0x8100000c00a4) = 0) AND (((S.flags_u & 0x400000000000) = 0) or (S.psfmagerr_u <= 0.2)) AND (((S.flags_u & 0x100000000000) = 0) or (S.flags_u & 0x1000) = 0) THEN S.psfMag_u ELSE null END AS psfMag_u,
CASE WHEN ((S.flags_u & 0x10000000) != 0) AND ((S.flags_u & 0x8100000c00a4) = 0) AND (((S.flags_u & 0x400000000000) = 0) or (S.psfmagerr_u <= 0.2)) AND (((S.flags_u & 0x100000000000) = 0) or (S.flags_u & 0x1000) = 0) THEN  S.psfMagErr_u ELSE null END AS psfMagErr_u,
CASE WHEN ((S.flags_g & 0x10000000) != 0) AND ((S.flags_g & 0x8100000c00a4) = 0) AND (((S.flags_g & 0x400000000000) = 0) or (S.psfmagerr_g <= 0.2)) AND (((S.flags_g & 0x100000000000) = 0) or (S.flags_g & 0x1000) = 0) THEN S.psfMag_g ELSE null END AS psfMag_g,
CASE WHEN ((S.flags_g & 0x10000000) != 0) AND ((S.flags_g & 0x8100000c00a4) = 0) AND (((S.flags_g & 0x400000000000) = 0) or (S.psfmagerr_g <= 0.2)) AND (((S.flags_g & 0x100000000000) = 0) or (S.flags_g & 0x1000) = 0) THEN  S.psfMagErr_g ELSE null END AS psfMagErr_g,
CASE WHEN ((S.flags_r & 0x10000000) != 0) AND ((S.flags_r & 0x8100000c00a4) = 0) AND (((S.flags_r & 0x400000000000) = 0) or (S.psfmagerr_r <= 0.2)) AND (((S.flags_r & 0x100000000000) = 0) or (S.flags_r & 0x1000) = 0) THEN S.psfMag_r ELSE null END AS psfMag_r,
CASE WHEN ((S.flags_r & 0x10000000) != 0) AND ((S.flags_r & 0x8100000c00a4) = 0) AND (((S.flags_r & 0x400000000000) = 0) or (S.psfmagerr_r <= 0.2)) AND (((S.flags_r & 0x100000000000) = 0) or (S.flags_r & 0x1000) = 0) THEN  S.psfMagErr_r ELSE null END AS psfMagErr_r,
CASE WHEN ((S.flags_i & 0x10000000) != 0) AND ((S.flags_i & 0x8100000c00a4) = 0) AND (((S.flags_i & 0x400000000000) = 0) or (S.psfmagerr_i <= 0.2)) AND (((S.flags_i & 0x100000000000) = 0) or (S.flags_i & 0x1000) = 0) THEN S.psfMag_i ELSE null END AS psfMag_i,
CASE WHEN ((S.flags_i & 0x10000000) != 0) AND ((S.flags_i & 0x8100000c00a4) = 0) AND (((S.flags_i & 0x400000000000) = 0) or (S.psfmagerr_i <= 0.2)) AND (((S.flags_i & 0x100000000000) = 0) or (S.flags_i & 0x1000) = 0) THEN  S.psfMagErr_i ELSE null END AS psfMagErr_i,
CASE WHEN ((S.flags_z & 0x10000000) != 0) AND ((S.flags_z & 0x8100000c00a4) = 0) AND (((S.flags_z & 0x400000000000) = 0) or (S.psfmagerr_z <= 0.2)) AND (((S.flags_z & 0x100000000000) = 0) or (S.flags_z & 0x1000) = 0) THEN S.psfMag_z ELSE null END AS psfMag_z,
CASE WHEN ((S.flags_z & 0x10000000) != 0) AND ((S.flags_z & 0x8100000c00a4) = 0) AND (((S.flags_z & 0x400000000000) = 0) or (S.psfmagerr_z <= 0.2)) AND (((S.flags_z & 0x100000000000) = 0) or (S.flags_z & 0x1000) = 0) THEN  S.psfMagErr_z ELSE null END AS psfMagErr_z
INTO mydb.OutputFile FROM Star as S
INNER JOIN MyDB.InputFile AS c ON S.objID = c.sdss_id
\end{spverbatim}

We also require $|G - g_{\rm{SDSS}} + 0.55| \leq 0.40$, to reduce the
number of spurious matches.

\item
Pan-STARRS1 $grizY$ photometry and associated uncertainties
\citep{2012ApJ...750...99T,2020ApJS..251....7F}, for sources as delineated
by the PS1 identifiers from the Gaia archive, are downloaded using
CASJobs (\url{https://mastweb.stsci.edu/ps1casjobs/default.aspx})using
the following query:
\begin{spverbatim}
SELECT a.objID,
CASE WHEN ((a.gQfPerfect > 0.85) AND (a.gMeanPSFMag > 14.5)) THEN a.gMeanPSFMag ELSE null END AS gMeanPSFMag,
CASE WHEN ((a.gQfPerfect > 0.85) AND (a.gMeanPSFMag > 14.5)) THEN a.gMeanPSFMagErr ELSE null END AS gMeanPSFMagErr,
CASE WHEN ((a.rQfPerfect > 0.85) AND (a.rMeanPSFMag > 15.0)) THEN a.rMeanPSFMag ELSE null END AS rMeanPSFMag,
CASE WHEN ((a.rQfPerfect > 0.85) AND (a.rMeanPSFMag > 15.0)) THEN a.rMeanPSFMagErr ELSE null END AS rMeanPSFMagErr,
CASE WHEN ((a.iQfPerfect > 0.85) AND (a.iMeanPSFMag > 15.0)) THEN a.iMeanPSFMag ELSE null END AS iMeanPSFMag,
CASE WHEN ((a.iQfPerfect > 0.85) AND (a.iMeanPSFMag > 15.0)) THEN a.iMeanPSFMagErr ELSE null END AS iMeanPSFMagErr,
CASE WHEN ((a.zQfPerfect > 0.85) AND (a.zMeanPSFMag > 14.0)) THEN a.zMeanPSFMag ELSE null END AS zMeanPSFMag,
CASE WHEN ((a.zQfPerfect > 0.85) AND (a.zMeanPSFMag > 14.0)) THEN a.zMeanPSFMagErr ELSE null END AS zMeanPSFMagErr,
CASE WHEN ((a.yQfPerfect > 0.85) AND (a.yMeanPSFMag > 13.0)) THEN a.yMeanPSFMag ELSE null END AS yMeanPSFMag,
CASE WHEN ((a.yQfPerfect > 0.85) AND (a.yMeanPSFMag > 13.0)) THEN a.yMeanPSFMagErr ELSE null END AS yMeanPSFMagErr
INTO mydb.OutputFile
FROM PanSTARRS_DR2.MeanObjectView AS a
INNER JOIN MyDB.InputFile AS b ON a.objid = b.panstarrs1_id
INNER JOIN (
   SELECT DISTINCT d.objid
   FROM PanSTARRS_DR2.StackObjectAttributes AS d
   INNER JOIN MyDB.InputFile AS e ON d.objid = e.panstarrs1_id
   WHERE d.primaryDetection > 0
) c ON a.objid = c.objid
WHERE a.nDetections > 5
AND (a.rMeanPSFMag - a.rMeanKronMag < 0.05)
\end{spverbatim}
Here, the values $14.5, 15.0, 15.0, 14.0, 13.0$ represent thresholds that
we use as precautions against saturation artefacts. We also require $|G -
g_{\rm{PS}} + 0.50| \leq 0.30$, to reduce the number of spurious matches.
\item
2MASS $JHK_{s}$ photometry and associated uncertainties
\citep{2006AJ....131.1163S} are downloaded from the NASA/IPAC Infrared
Science Archive (\url{https://irsa.ipac.caltech.edu/frontpage/}). For
each source, we require \code{use\_src=1}, for each band we require
\code{(ph\_qual='A') or (rd\_flg=\{'1','3'\})}, and we require a
1$\arcsec$ match between the 2MASS astrometry and that from the ra2000
and dec2000 astrometry computed from Gaia DR3 astrometry.
\item
WISE $W_{1}W_{2}$ photometry and associated uncertainties
\citep{2010AJ....140.1868W} are downloaded from the NASA/IPAC Infrared
Science Archive. For brighter sources ($8 \geq W_{1} \geq 2,\, 7 \geq
W_{2} \geq 1.5$) we use the WISE All-Sky photometric catalog, whereas
for fainter ($W_{1} \geq 8,\, W_{2} \geq 7$) sources we use the AllWISE
catalog. For each bandpass, we require \code{(ph\_qual=\{'A','B'\})},
\code{(ext\_flag=0)}, and \code{cc\_flags=\{'0','d','p','h','o'\}}.
We also download the $W_{3}$ and $W_{4}$ photometry and associated
uncertainties, but they do not contribute to our analysis.
\end{enumerate}

\section{Spectroscopic Comparison Data Catalog Construction and
Quality Flags}

\begin{enumerate}
\item
Gaia spectroscopic parameters from the high-resolution channel
are downloaded from the Gaia data archive,  specifically from
the \texttt{gaiadr3.astrophysical\_parameters} catalog. For the
parameter \texttt{flags\_gspspec}, which has an injective mapping onto
\texttt{gaiadr3.source\_id}, we require that the first six parameters of
the string be equal to `0', the seventh character be one of \{'0','1'\},
the eighth, eleventh, and twelfth characters be equal to `0'. There are
no constraints on the remaining characters.
\item
APOGEE spectroscopic parameters are downloaded from the DR17.aspcapStar
catalog on CASJobs, subject to the criteria \texttt{(a.aspcapflag \&
261033871) = 0}, as well as  \texttt{((a.teff\_flag \& 87903) = 0) AND
(a.teff > -9999))}, and similarly for the parameters \texttt{logg},
\texttt{m\_h}, and \texttt{fe\_h}. We used the following query:
\begin{spverbatim}
SELECT a.apogee_id, b.gaiaedr3_source_id AS dr3_source_id,
CASE WHEN (((a.teff_flag & 87903) = 0) AND (a.teff > -9999)) THEN a.teff ELSE null END AS teff,
CASE WHEN (((a.teff_flag & 87903) = 0) AND (a.teff > -9999)) THEN a.teff_err ELSE null END AS teff_err,
CASE WHEN (((a.logg_flag & 87903) = 0) AND (a.logg > -9999)) THEN a.logg ELSE null END AS logg,
CASE WHEN (((a.logg_flag & 87903) = 0) AND (a.logg > -9999)) THEN a.logg_err ELSE null END AS logg_err,
CASE WHEN (((a.m_h_flag & 87903) = 0) AND (a.m_h > -9999)) THEN a.m_h ELSE null END AS m_h,
CASE WHEN (((a.m_h_flag & 87903) = 0) AND (a.m_h > -9999)) THEN a.m_h_err ELSE null END AS m_h_err,
CASE WHEN (((a.fe_h_flag & 87903) = 0) AND (a.fe_h > -9999)) THEN fe_h ELSE null END AS fe_h,
CASE WHEN (((a.fe_h_flag & 87903) = 0) AND (a.fe_h > -9999)) THEN fe_h_err ELSE null END AS fe_h_err into mydb.OutputFile from DR17.aspcapStar a
INNER JOIN DR17.apogeeStar b ON a.apstar_id = b.apstar_id
WHERE (a.aspcapflag & 261033871) = 0
\end{spverbatim}
\item
The main GALAH DR3 spectroscopic catalog,
\texttt{GALAH\_DR3\_main\_allstar\_v2.fits}, is downloaded from the
GALAH webpage (\url{https://www.galah-survey.org/}). We require that
\texttt{snr\_c3\_iraf > 30}, \texttt{flag\_sp = 0}, and \texttt{flag\_fe\_h
= 0}.
\item
The LAMOST DR7 data selection are as described by
\citet{2021RNAAS...5...24H}.
\end{enumerate}

\end{document}